\documentclass[reprint,superscriptaddress,amsmath,amssymb,aps,longbibliography,floatfix]{revtex4-2}

\usepackage{graphicx}
\usepackage[normalem]{ulem}
\usepackage[svgnames]{xcolor}
\usepackage{bm}
\usepackage{multirow}
\usepackage{float}
\usepackage{titlesec}
\usepackage[columnwise]{lineno}
\usepackage{titlesec}
\usepackage{gensymb}
\usepackage{textcomp}

\begin{document}
\title{Phonon drag thermal Hall effect in metallic strontium titanate}

\author{Shan Jiang}
\affiliation{Laboratoire de Physique et d'Étude des Matériaux\\ (ESPCI Paris - CNRS - Sorbonne Universit\'e), PSL University, 75005 Paris, France}

\author{Xiaokang Li}
\affiliation{Laboratoire de Physique et d'Étude des Matériaux\\ (ESPCI Paris - CNRS - Sorbonne Universit\'e), PSL University, 75005 Paris, France}
\thanks{Present address: Wuhan National High Magnetic Field Center and School of Physics, Huazhong University of Science and Technology, Wuhan 430074, China}
\author{Beno\^it Fauqu\'e}
\affiliation{JEIP, USR 3573 CNRS, Coll\`ege de France, PSL University, 75231 Paris Cedex 05, France}
\author{Kamran Behnia} 
\affiliation{Laboratoire de Physique et d'Étude des Matériaux\\ (ESPCI Paris - CNRS - Sorbonne Universit\'e), PSL University, 75005 Paris, France}

\date{\today}
\begin{abstract}
SrTiO$_3$, a quantum paralectric, displays a detectable phonon thermal Hall effect (THE). Here we show that the amplitude of THE is extremely sensitive to stoichiometry. It drastically decreases upon substitution of a tiny fraction of Sr atoms with Ca, which stabilizes the ferroelectric order. It drastically increases by an even lower density of oxygen vacancies, which turn the system to a dilute metal. The enhancement in the metallic state exceeds by far the sum of the electronic and the  phononic contributions. We explain this observation as an outcome of three features: i)  heat is mostly transported by phonons;  ii) the electronic Hall angle is extremely large; and iii) there is substantial momentum exchange between electrons and phonons. Starting from Herring's picture of phonon drag, we arrive to a quantitative account of the enhanced THE. Thus, phonon drag, hitherto detected as an amplifier of thermoelectric coefficients, can generate a purely thermal transverse response in a dilute metal with a large Hall angle. Our results reveal a hitherto unknown consequence of momentum-conserving collisions between electrons and phonons.
\end{abstract}

\maketitle

\section*{Introduction}
The observation of thermal Hall effect (THE) in a variety of insulators \cite{Sugii2017,Kasahara2018,Grissonnanche2019, Hirokane2019,Li2020,bruin2021,lefrancois2021} has attracted much recent attention. A transverse thermal gradient produced by a longitudinal flow of neutral carriers of heat requires a microscopic mechanism other than the Lorentz force. In strontium titanate, a non-magnetic insulator \cite{Collignon2019}, the phononic origin of the THE \cite{Li2020,Sim2021} is uncontested.  Theoretical scenarios \cite{Shin2012, Chen2020, Sun2021b, Flebus2021,Guo2021, Zabalo2021,Bhalla_2021} have been proposed to explain how phonons can generate a transverse thermal gradient on top of the longitudinal one.

Ferroelectric transition is aborted by quantum critical fluctuations \cite{Rowley2014} in strontium titanate. The ground state of this quantum paraelectric \cite{Muller1979} is unusually sensitive to the presence of extrinsic atoms  \cite{Bednorz1984,Lemanov1996,Itoh1999}. Sr$_{1-x}$Ca$_x$TiO$_3$ is ferroelectric for $x>0.002$ \cite{Bednorz1984,Carpenter2006}. Introducing a tiny amount of oxygen vacancies, on the other hand, makes the system metallic. With a carrier density of the order of 10$^{17}cm^{-3}$: SrTiO$_{3-\delta}$ is a dilute metal with a sharp Fermi surface and a superconducting ground state \cite{Lin2013}. Double substitution leads to Sr$_{1-x}$Ca$_x$TiO$_{3-\delta}$, a polar metal \cite{Wang2019,Engelmayer2019}, where superconductivity is boosted \cite{Rischau2017}. Here, we present a study of THE in Ca-substituted [insulating] and in oxygen-reduced [metallic] samples of strontium titanate. We find that the amplitude of the signal is significantly modified in both cases, but in opposite directions. In the case of Ca substitution, we find that THE is drastically reduced, confirming that stabilization of the ferroelectric order is detrimental to THE \cite{Sim2021}. On the other hand, in SrTiO$_{3-\delta}$, the amplitude of THE not only exceeds what is found in the undoped insulator, but is also much larger than the sum of the expected electronic contribution and the phononic one. We argue that this surprising result can be understood by invoking the drag \cite{Herring1954,Gurevich1989} between electrons and phonons. In our temperature range of interest, heat is almost exclusively carried by phonons ($\kappa^{ph} \gg \kappa^{e} $) in this dilute metal. On the other hand, the phonon Hall angle is small ($\kappa^{ph}_{xy}< 10^{-3} \kappa^{ph}_{xx}$), but not the electronic Hall angle, which exceeds unity ($\sigma_{xy} > \sigma_{xx} $  at $\approx  2T$). In this context, momentum-conserving collisions between electrons and phonons can generate a transverse temperature gradient, because longitudinal momentum transferred to the electron bath will have a transverse counterpart, which ultimately generates a transverse energy flow. Starting from Herring's picture of phonon drag \cite{Herring1954}, we will show that this scenario yields a quantitative account of our experimental result. 

Such a conclusion implies that phonon drag, hitherto known as an amplifier of the thermoelectric response \cite{MacDonald}, can have a purely thermal signature, mostly in the transverse channel.  In this context, our result identifies a  previously unknown consequence of quasi-particle hydrodynamics \cite{gurzhi1968}. Momentum-conserving collisions between phonons and electrons \cite{jaoui2022} are a subject of growing interest in strongly-coupled electron-phonon systems \cite{levchenko2020,lucas2021}. Momentum-conserving phonon-phonon collisions are also known to play a role in setting the amplitude of longitudinal thermal conductivity in Callaway's model \cite{Callaway1959}. 
\begin{figure*}
\centering
\includegraphics[width=.75\linewidth]{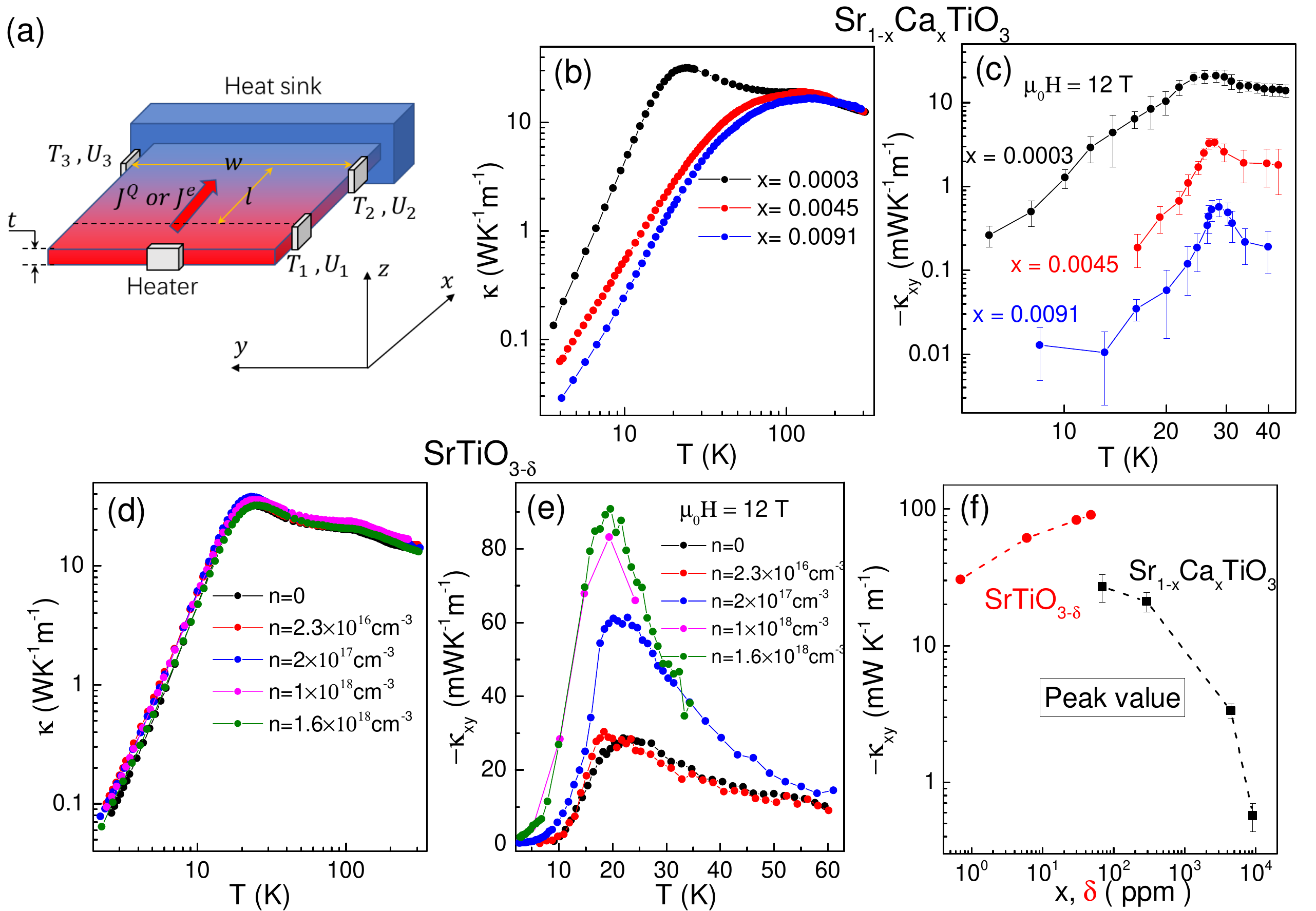}
\caption{\textbf{Evolution of longitudinal and transverse thermal conductivity with Ca substitution and oxygen reduction: }(a) Setup for measuring longitudinal and transverse thermal, thermoelectric and electric coefficients (the yellow pads are gold electrodes). We  measured three local temperatures T$_1$, T$_2$ and T$_3$ and three local voltages U$_1$, U$_2$ and U$_3$. This allowed us to measure longitudinal and transverse temperature gradients and electric fields caused by a longitudinal heat (J$^Q$) or electric (J$^e$) current.  Temperature dependence of the longitudinal, $\kappa_{xx}$ (b), and transverse, $\kappa_{xy}$  (c),  thermal conductivity in Sr$_{1-x}$Ca$_x$TiO$_3$. The 20 K peak in $\kappa_{xx}$ is wiped out. $\kappa_{xy}$ decreases by almost two orders of magnitude. (d) Temperature dependence of $\kappa_{xx}$ in SrTiO$_{3-\delta}$, almost unaffected by doping. (e) Temperature dependence of $\kappa_{xy}$ in SrTiO$_{3-\delta}$. The amplitude enhances with oxygen substitution. (f) Amplitude of $\kappa_{xy}$ peak as a function of concentration of Ca atoms (x) and O vacancies ($\delta$) in parts per million (ppm). }
\label{fig: evolution}
\end{figure*}

\section*{Results}
Our set-up (See Fig. \ref{fig: evolution} a) was designed to determine the electrical, thermoelectric and thermal transport coefficients of a single crystal during a single run. In insulating Ca-doped SrTiO$_3$ single crystals, we measured longitudinal and transverse thermal conductivities.  In the case of metallic SrTiO$_{3-\delta}$ single crystals, we measured longitudinal and tranverse conductivities in thermal and electrical channels, as  well as the Seebeck and the Nernst coefficients. In the latter case, by performing two sets of experiments one with a zero electrical current and another with a zero thermal current, we checked that the transport coefficients respect Onsager reciprocity.

\subsection*{Calcium substitution} The evolution of longitudinal and transverse thermal conductivity  with Ca substitution is shown in  Fig. \ref{fig: evolution} b,c. One can clearly see that both are affected by Ca substitution. The decrease in $\kappa_{xx}$  can be attributed to the random distribution of Ca atoms that introduce additional scattering. As  in the case of Nb substitution \cite{Martlelli2018}, introducing less than a percent concentration of extrinsic atoms is sufficient to wipe out the 20 K peak of the thermal conductivity. The decrease in  $\kappa_{xy}$ is even more drastic. Substituting  a tiny fraction (x=0.002) of Sr atoms with Ca is sufficient for stabilizing a long-range ferroelectric order \cite{Bednorz1984}. This substitution eventually leads to a fifty-fold decrease in the magnitude of $\kappa_{xy}$. 

A previous study on the effect of $^{18}$O substitution on the thermal Hall effect in strontium titanate \cite{Sim2021} found  a decrease of comparable magnitude. Substituting Sr by Ca \cite{Rischau2017,Wang2019} and substituting $^{16}$O with $^{18}$O \cite{Itoh1999,Rischau2022} both stabilize the ferroelectric order, modify the superconducting dome and generate a polar metal. In both cases, the large $\kappa_{xy}$ of the quantum paraelectric solid is drastically suppressed with the stabilization of the ferroelectric order. This suggests a link between the amplitude of $\kappa_{xy}$  and the presence of ferroelectric fluctuations \cite{Chen2020}. Note that in $^{18}$O-enriched strontium titanate \cite{Sim2021},  most $^{16}$O atoms were substituted and the 20 K peak in $\kappa_{xx}$ is still present, in contrast to what is seen here. Yet, $\kappa_{xy}$ was similarly damped, ruling out that $\kappa_{xy}$ is simply more sensitive to disorder.

\subsection*{Oxygen reduction} We investigated the effect of oxygen vacancies on THE by studying thermal transport in SrTiO$_{3-\delta}$. Fig. \ref{fig: evolution}d,e shows how the temperature dependence of longitudinal and transverse thermal conductivity  in SrTiO$_{3-\delta}$. The magnitude of longitudinal $\kappa_{xx}$ barely changes with oxygen reduction. Assuming that each vacancy introduces two electrons, the concentration of vacancies for n $= 1.6 \times 10^{18} cm^{-3}$ (determined by measuring its Hall resitivity) is only $\delta=4.8\times 10^{-5}$ per formula unit, much lower than the lowest amount of Ca substitution (x$=4.5\times 10^{-3} $). By measuring the  electric conductivity and using the Wiedemann-Franz (WF) law, we estimated the amplitude of the electronic heat conductivity.  $\kappa^{e}$  remains less than $10^{-2}$ of the total thermal conductivity. Therefore, thermal conductivity near the peak is not significantly reduced by disorder or enhanced by the addition of a finite electronic contribution. In contrast, $\kappa_{xy}$ significantly increases with increasing $n$.  Let us compare this enhancement with the expected electronic contribution.

\begin{figure}[t]
\centering
\includegraphics[width=.8\linewidth]{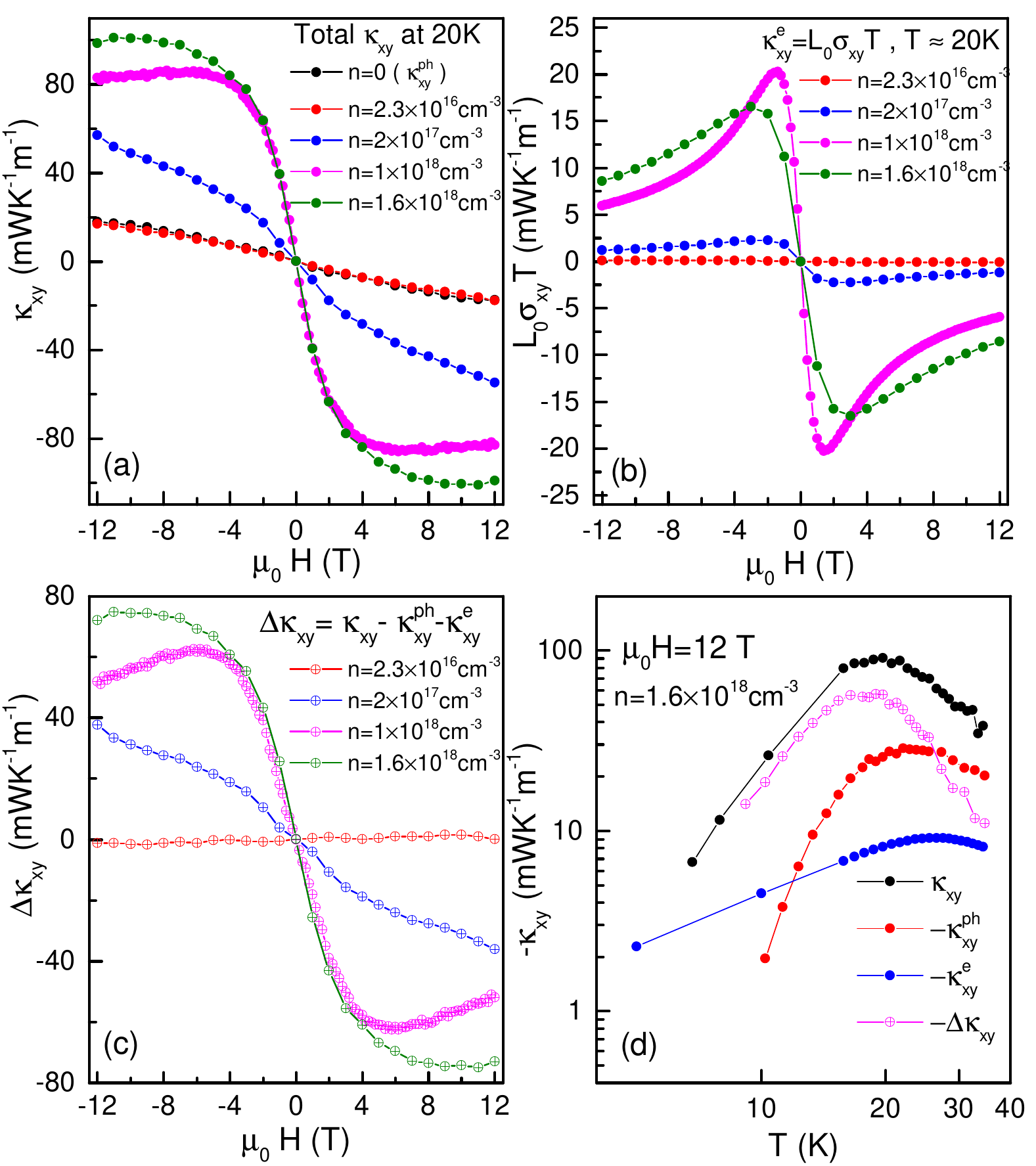}
    \caption{\textbf{Three components of the thermal Hall effect in SrTiO$_{3-\delta}$ :} (a) $\kappa_{xy}$ as a function of magnetic field  in samples with different carrier densities at 20 K. With growing metallicity, the amplitude of $\kappa_{xy}$ increases and its field dependence becomes less linear. (b) Field dependence of the electrical Hall conductivity, $\sigma_{xy}$, multiplied by $L_0=\frac{\pi^2}{3}\frac{k_B^2}{e^2}$, and the temperature, $T$, in the metallic samples. (c) Field dependence of the differential  $\Delta\kappa_{xy}= \kappa_{xy}-\kappa_{xy}(n=0)-L_0\sigma_{xy}T$, which signals the presence of a third term in addition to the purely electronic and the purely phononic terms. (d) The evolution of $\kappa_{xy}$ and its three components as a function of temperature. Below 25 K, $\Delta\kappa_{xy}$ is the largest component.}
\label{fig: field dependence}
\end{figure}

Fig. \ref{fig: field dependence}a compares the field dependence of  $\kappa_{xy}$ in different samples at 20 K (i.e. near its peak). In the sample with the lowest carrier concentration ($n=2.3\times10^{16}cm^{-3}$), the amplitude of $\kappa_{xy}$ is almost identical to the insulating sample and remains linear in magnetic field. In samples with higher carrier concentration, the amplitude of $\kappa_{xy}$ is larger and shows a gradual trend towards high-field saturation. Fig. \ref{fig: field dependence}b shows the electronic  contribution to THE (estimated from the measured electric Hall conductivity of the metallic samples: $\kappa^e_{xy}=L_0\sigma_{xy}T$). It shows a field-induced saturation similar to what is observed in $\kappa_{xy}$. However, subtracting the purely phononic component (taken to be equal to what is observed in the insulator) and the purely electronic component (estimated from the WF law) from the total $\kappa_{xy}$ leaves us with an additional  $\Delta\kappa_{xy}= \kappa_{xy}-\kappa^{ph}_{xy}-\kappa^{e}_{xy}$. Fig. \ref{fig: field dependence}d shows the temperature dependence of the different components of THE in the sample with the highest carrier concentration ($n=1.6\times10^{18}cm^{-3}$). One can see that $\Delta\kappa_{xy}$ exceeds  $\kappa^{ph}_{xy}$ and is several times larger than $\kappa^{e}_{xy}$ in most of the temperature range.  Note also that the temperature dependence of $\Delta\kappa_{xy}$ is significantly different from $\kappa^{e}_{xy}$. 

We shall keep in mind that at finite temperature the WF law is not strictly valid. However, the finite-temperature correction does not modify the order of magnitude of the expected electronic thermal conductivity. Moreover, since inelastic scattering damps thermal transport more than the electrical transport, the expected correction  to the Lorenz ratio is downward. In copper, such a downward  deviation have been observed in the transverse channel of heat and charge transport as well as in the longitudinal ones \cite{Zhang2000}. Thus,  $L_0\sigma_{xy}T$ gives an upper bound to the expected  $\kappa^{e}_{xy}$ and we can safely conclude that enhanced THE is not simply due to the introduction of mobile electrons. 

\subsection*{The thermoelectric correction to the thermal conductivity} Before discussing the origin of this additional component, we need to distinguish between two thermal conductivities  \cite{Behnia2015b}. The first one is defined by the Fourier equation:  
\begin{equation}
  \Vec{J^Q} = -\overline{\kappa} \Vec{\nabla T} 
\end{equation}

In presence of thermoelectric phenomena, the transport equations become:
\begin{eqnarray}
  \Vec{J^e} = \overline{\sigma} \Vec{E} -\overline{\alpha} \Vec{\nabla T} \\
  \Vec{J^Q} = \overline{\alpha} T \Vec{E} -\overline{\kappa^{\prime}} \Vec{\nabla T} 
\end{eqnarray}
where $\overline{\sigma}$ and $\overline{\alpha}$ are 
the electric and thermoelectric conductivity tensors. Now if the charge current is kept equal to zero ($\Vec{J^e}=\Vec{0}$), the combination of the two equations would yield:

\begin{equation}
  \Vec{J^Q} = (\overline{\alpha}\,\overline{\rho}\,\overline{\alpha}\,T-\overline{\kappa^{\prime}}) \Vec{\nabla T} 
  \label{TE-correction}
\end{equation}

The resistivity tensor is simply the inverse of the conductivity tensor: $\overline{\rho}=\overline{\sigma}^{-1}$. Note that the true Onsager coefficient is $\overline{\kappa^{\prime}}$  and not $\overline{\kappa}$. Practically, this distinction matters only when the first term on the right hand of Eq. \ref{TE-correction} is not negligible compared to the second term, which happens when the thermoelectric figure of merit is sizeable \cite{Behnia2015b}. 

We have quantified this difference by measuring longitudinal and lateral temperature differences and electric fields in two distinct  experiments. In the first, a finite  $\Vec{J^Q}$ was applied and the  $\Vec{J^e}$ was kept equal to zero.  In the second,  a finite    $\Vec{J^e}$ was injected without $\Vec{J^Q}$. This led us to quantify the diagonal and off-diagonal components of the two conductivities, $\overline{\kappa}$ and $\overline{\kappa^\prime}$. We also checked that the  data respect Onsager reciprocity, which implies a unique thermoelectric tensor in equations (2) and (3), (See the supplement for details \cite{SM}).

\begin{figure*}
\centering
\includegraphics[width=.8\linewidth]{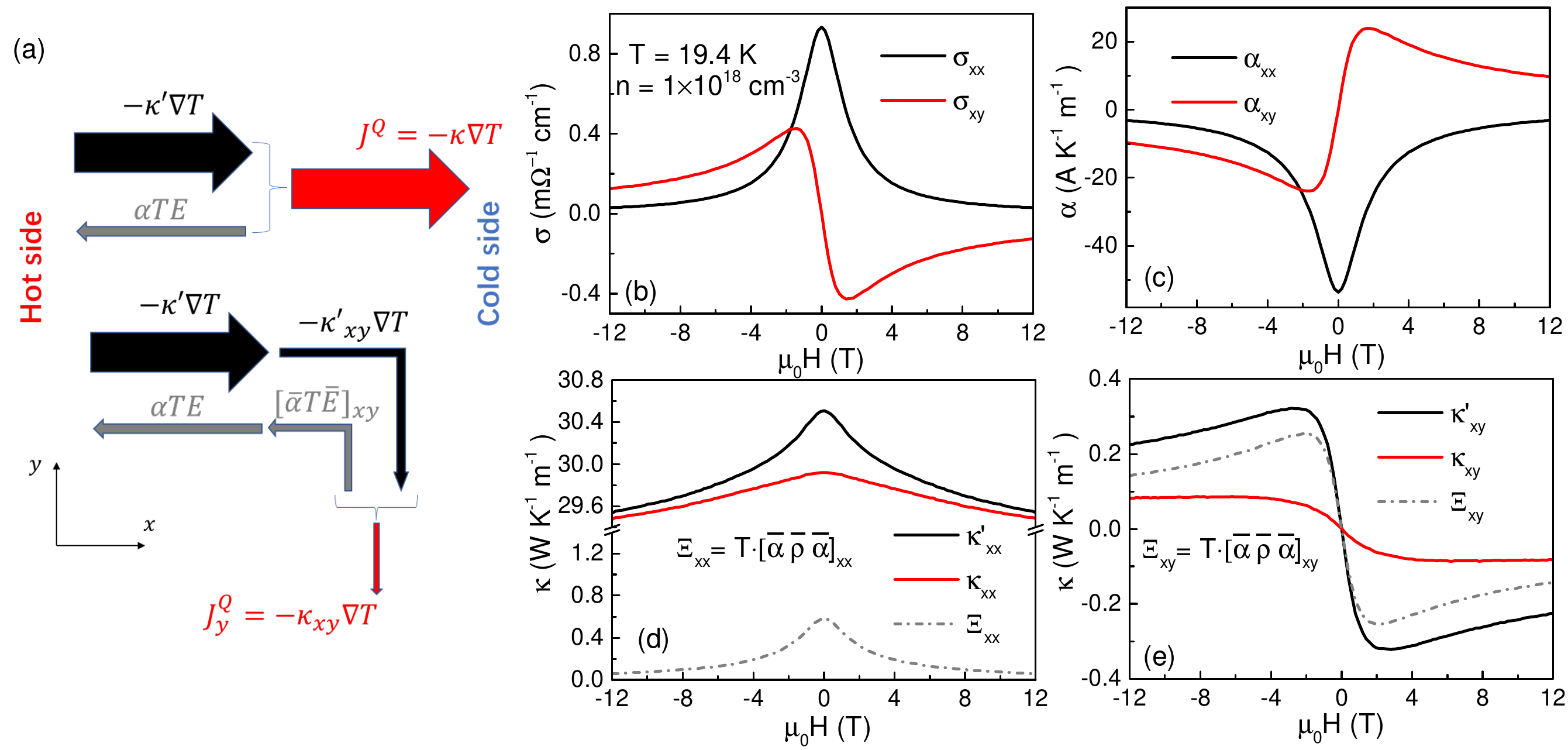} 
\caption{\textbf{The two components of heat flux, :} (a) The heat current density, $\Vec{J^Q}$, consists of two terms in both longitudinal and transverse channels. The first component ($- \overline{\kappa^\prime} \Vec{\nabla T}$) represents the flow of entropy  without particle flow. The second ($\overline{\alpha} T$ ) is the flow of entropy thanks to particle flow. The difference between $\kappa$ and $\kappa^\prime$ is significant when the second term is not negligible. In our case, this happens for the transverse channel. The width of arrows schematize the weight of different components. (b) The longitudinal and the transverse electric conductivity as a function of field at 20 K for SrTiO$_{3-\delta}$ (n=$1\times10^{18}cm^{-3}$).(c) Same for longitudinal and transverse thermoelectric conductivity . $\overline{\alpha} = \overline{\sigma}\,\overline{S}$. Note that the off-diagonal components rapidly become as large as the diagonal components. (d) $\kappa_{xx}$ and $\kappa^\prime_{xx}$ as function of magnetic field. The difference is small. (e) $\kappa_{xy}$, $\kappa^\prime_{xy}$  as function of magnetic field. The difference is significant. Also shown in the panels are the diagonal and the off-diagonal components of the $\overline{\Xi}=\overline{\alpha}\,\overline{\rho}\, \overline{\alpha}\,T$ tensor, which quantifies the correction. }
\label{fig: kappa'}
\end{figure*}

We found that the transverse thermal flow (but not the longitudinal one) is drastically affected by the particle-driven flow of entropy represented by the thermoelectric term (See Fig \ref{fig: kappa'} a). In other words, $\kappa_{xx} \simeq \kappa^\prime_{xx}$, but  $\kappa_{xy}$ and  $\kappa^\prime_{xy}$ are significantly different. The reason is the large Hall angle in the electric and the thermoelectric response. As seen in  Fig \ref{fig: kappa'} b, c, the diagonal and off-diagonal components of $\overline{\sigma}$ and $\overline{\alpha}$ are of the same order of magnitude. In contrast, the diagonal component of the thermal conductivity tensor (Fig \ref{fig: kappa'} d) is orders of magnitude larger than the off-diagonal one (Fig \ref{fig: kappa'} e). As a consequence, while the difference between $\kappa_{xx}$ and $\kappa^\prime_{xx}$ is of the order of percent (Fig \ref{fig: kappa'} d),  $\kappa^\prime_{xy}$ is 2 times larger than $\kappa_{xy}$ (Fig \ref{fig: kappa'} e).  

Thus, the corrected transverse thermal conductivity is larger than the measured one, which is itself larger than the expected one. The large electric and thermoelectric Hall angles which led to this correction is a key ingredient of the solution to this puzzle. Let us now consider the others.

\begin{figure}
\centering
\includegraphics[width=.8\linewidth]{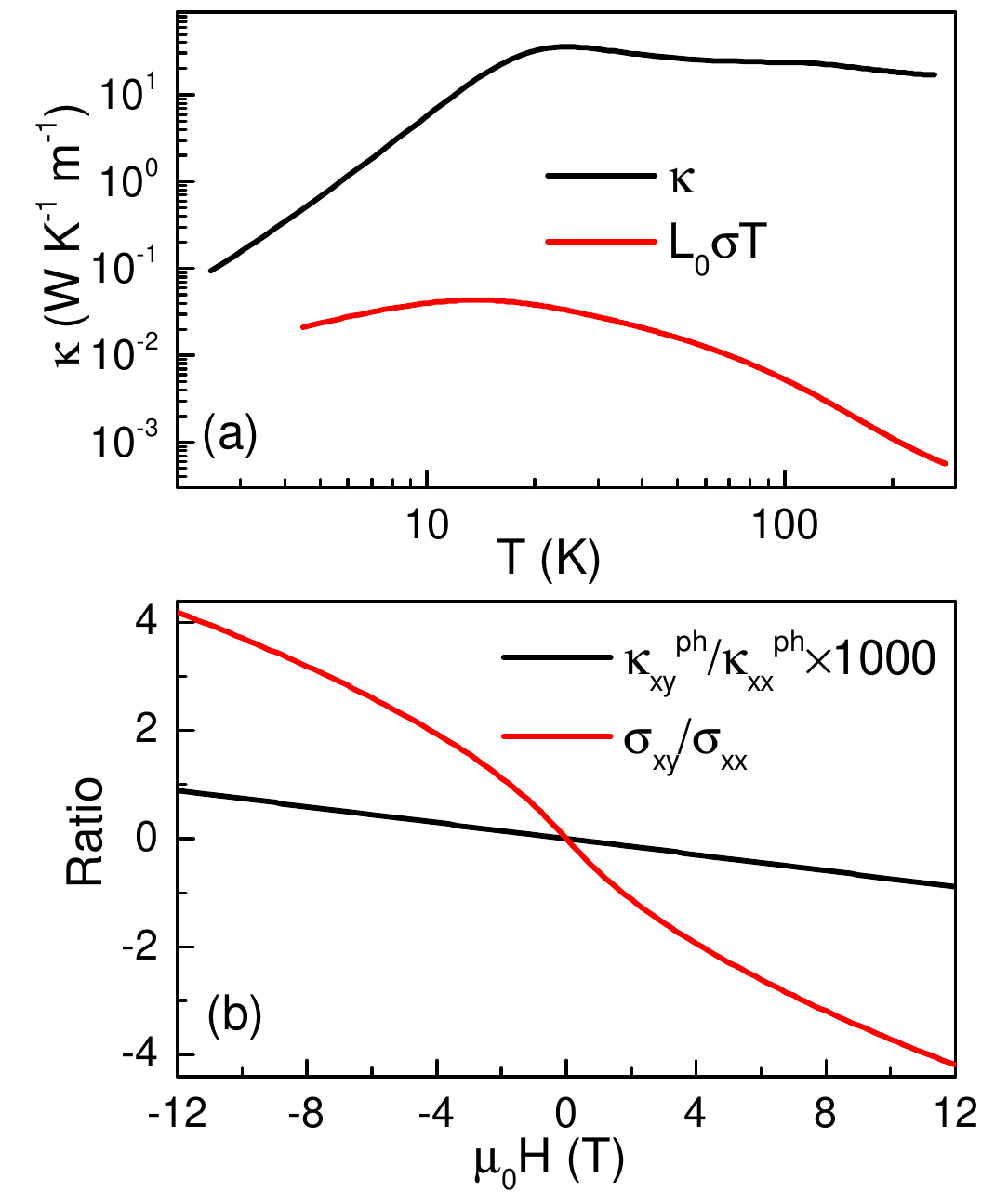} 
\caption{\textbf{Longitudinal thermal conductivity and Hall angles}  (a) Thermal conductivity ($\kappa_{xx}$) of SrTiO$_{3-\delta}$ ($n=1\times10^{18} cm^{-3}$) as function of temperature and its electronic component ($L_{0}\sigma T)$. Phonons are by far the dominant carriers of heat. (b) The phonon thermal and the electronic Hall angles as the function of field at T= 20 K. The former is three orders of magnitude smaller than the latter. }
\label{fig: properties}
\end{figure}

\section*{Discussion}
\subsection*{The three ingredients of the scenario }
Our scenario invokes three different features of lightly doped strontium titanate. The first is that momentum exchange between phonons and electrons is frequent. Since both the Fermi radius and the phonon thermal wavelength  (at our temperature range of interest) are much smaller than the width of the Brillouin zone, such collisions are not Umklapp and conserve momentum. The second feature is that heat is mainly carried by phonons (and not by electrons), $\kappa_{xx}^e\ll \kappa_{xx}^{ph}$, see Fig \ref{fig: properties} a).  The third is that the Hall angle of electrons exceeds by far the (thermal) Hall angle of phonons. As one can see in Fig \ref{fig: properties} b, the phonon thermal Hall angle is 3 orders of magnitude smaller than the electric Hall angle. Lightly-doped strontium titanate is a dilute metal with highly mobile carriers. A moderate magnetic field puts the system in the high field limit, where $\mu B \gg 1$ \cite{Collignon2021}. The combination of these three features generates an additional thermal Hall response: The longitudinal energy flow, mostly carried by phonons, is accompanied by a flow of electrons and its unavoidable transverse counterpart, which ends up by triggering a phononic transverse flow.

Momentum exchange between electrons and phonons (in presence of heat flow and in absence of charge current) is known as phonon drag  \cite{Herring1954,Gurevich1989}. It is known to amplify the thermoelectric response, mostly in semiconductors \cite{Herring1954}, but also in metals \cite{MacDonald}. In the case of strontium titanate, previous studies \cite{Cain2013,Collignon2020} have shown that phonon drag causes a peak in the zero- field Seebeck coefficient around $\sim 20$ K. We confirmed the presence of such a peak in our samples (See the supplement \cite{SM}). Let us now quantify the expected contribution of phonon drag to transverse thermal transport conductivity. 

\subsection*{From Phonon drag to thermal Hall effect}
Phonons streaming from hot to cold exert a drag on the charge carriers. To quantify this effect, Herring considered an equivalent phenomenon: the enhancement in the Peltier coefficient of an isothermal sample caused by the drag exerted on phonons by the electric current \cite{Herring1954}. Assuming an approximate proportionality between heat current and crystal momentum, he found that a phonon drag Peltier effect, $\Pi_{drag}$, of either sign can arise : 
\begin{equation}
    \Pi_{drag}=\pm \frac{m^*v_s^2}{e} f\frac{\tau_p}{\tau_e}
    \label{Herring}
\end{equation}

Here, $m^*$ is the effective mass of electrons, $v_s$ is the sound velocity, $e$ is the fundamental charge, $\tau_p$ and $\tau_e$ are respectively, the phonon and the electron scattering times and  $0<f<1$ represents the fraction of collisions suffered by phonons, which leads to momentum exchange between the phonon bath and the electron bath. Using the Kelvin relation, the phonon drag component of the Seebeck coefficient becomes $\Pi_{drag}/T$. To derive Eq. \ref{Herring}, Herring put forward two arguments. First of all the energy  density flux, $J^Q$, can be approximated by the product of the crystal momentum per unit volume, $P$ and the square of sound velocity, $v_s$: 
\begin{equation}
    J^Q= P v_s^2
    \label{flux}
\end{equation}
The second argument is that the rate at which phonons receive crystal momentum from the electronic carriers is to be balanced with  the rate at which they lose it. Therefore: 
\begin{equation}
   \frac{P}{\tau_p}= \pm f n e E
    \label{balance}
\end{equation}

Here, $n$ is the carrier concentration, and $E$ is the electric field. The loss of crystal momentum out of the phonon bath is countered by what electrons introduce to this bath.  The parameter $f$ is a measure of efficiency of momentum flow between the phonon and electron baths. Herring invoked a 'hydraulic analogy' where momentum first flows from electrons to small $q$ phonons and then distributed among phonons and then eventually lost. In this analogy $f$ plays the role of a half-open valve. Eq.\ref{Herring} can be obtained from Eq.\ref{flux} and Eq.\ref{balance}, using the Drude link between the electric field and the electric current ($J^e=\frac{ne^2\tau_e}{m^*}E$). 

\begin{figure}
\centering
\includegraphics[width=.9\linewidth]{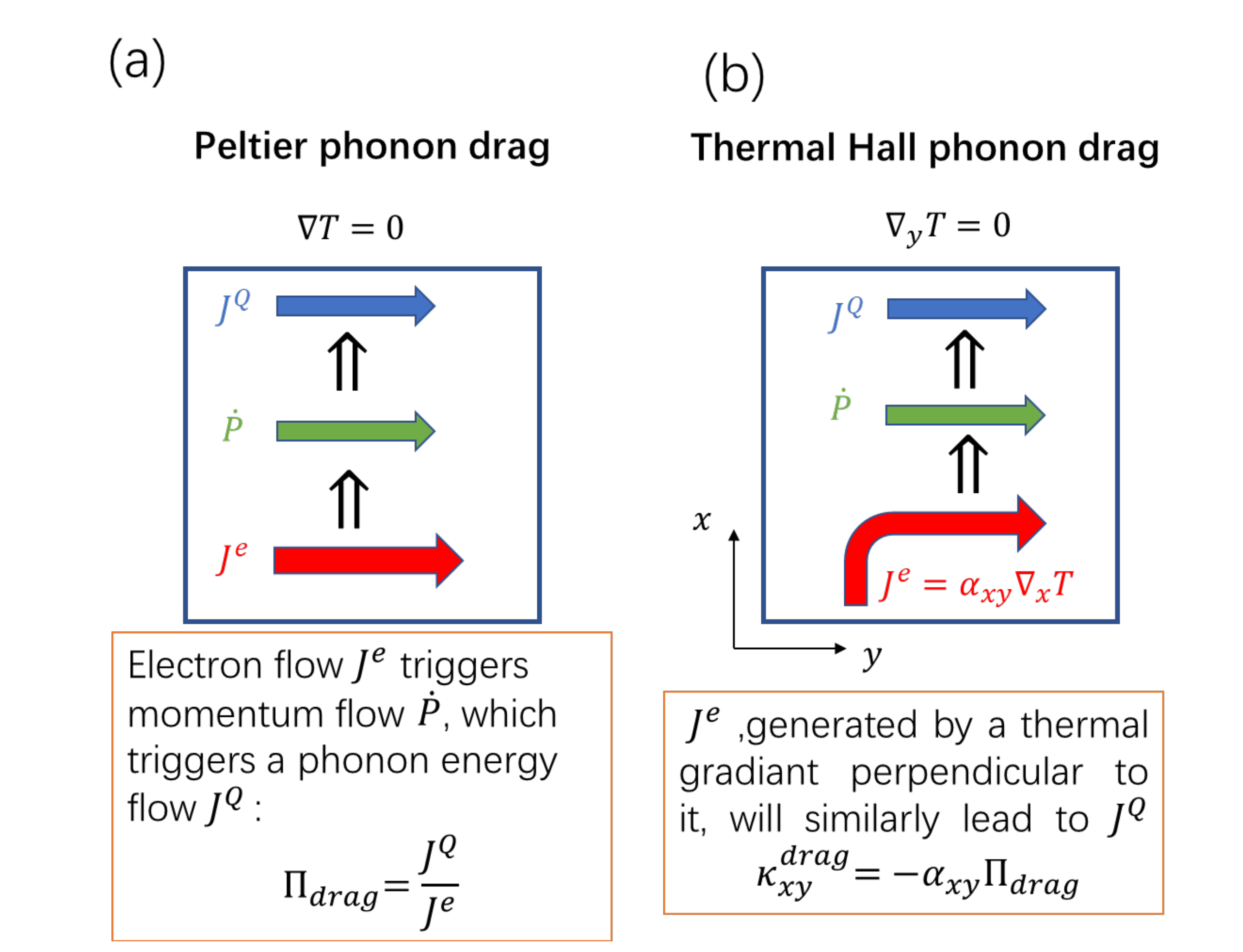}
\caption{\textbf{From phonon drag to thermal Hall conductivity:} (a) Herring's picture of phonon drag in an isothermal sample. Electronic momentum flow generates a phonon momentum flow which in turn leads to a phonon heat flow. The amplitude of this Peltier effect sets the amplitude of the phonon drag Seebeck effect. (b) Replacing the  electronic charge current with a finite transverse thermal gradient multiplied by off-diagonal thermoelectric conductivity quantifies the amplitude of the phonon drag thermal Hall effect. }
\label{fig: sketch}
\end{figure}

Let us now consider the twist brought by a large Hall angle to this picture. The Peltier phonon drag implies that an electric current can lead to a phonon energy flow (Fig. \ref{fig: sketch}a). A finite THE driven by phonon drag can be explained in the following way. A finite Nernst-Ettingshausen coefficient implies that a  transverse (longitudinal) thermal gradient will generate a longitudinal (transverse) electric current.Then, this electric current, following Herring's original picture, will generate a thermal current. The latter will be perpendicular to the thermal gradient (Fig. \ref{fig: sketch}b). Therefore, the overall magnitude of this phonon drag THE will be given by the product of the off-diagonal thermoelectric conductivity, $\alpha_{xy}$, and Herring's expression for Peltier phonon drag: 
\begin{equation}
    \kappa_{xy}(drag) = \alpha_{xy} \frac{m^*v_s^2}{e}f\frac{\tau_p}{\tau_e}
    \label{kappa_ep}
\end{equation}

Thus, the component of thermal Hall conductivity generated by mutual drag between electrons and phonons is proportional to the product of $\alpha_{xy}$, the ratio of phonon and electron scattering times $\frac{\tau_p}{\tau_e}$ and the efficiency of momentum transfer between the two baths, parametrized by $f$. Let us also note the presence of $mv_s^2$. This is the kinetic energy of an electron drifting with the velocity of sound, a less familiar energy scale emerging when electrons and phonons couple to each other \cite{Mousatov2021}.   

\begin{figure*}
\centering
\includegraphics[width=.8\linewidth]{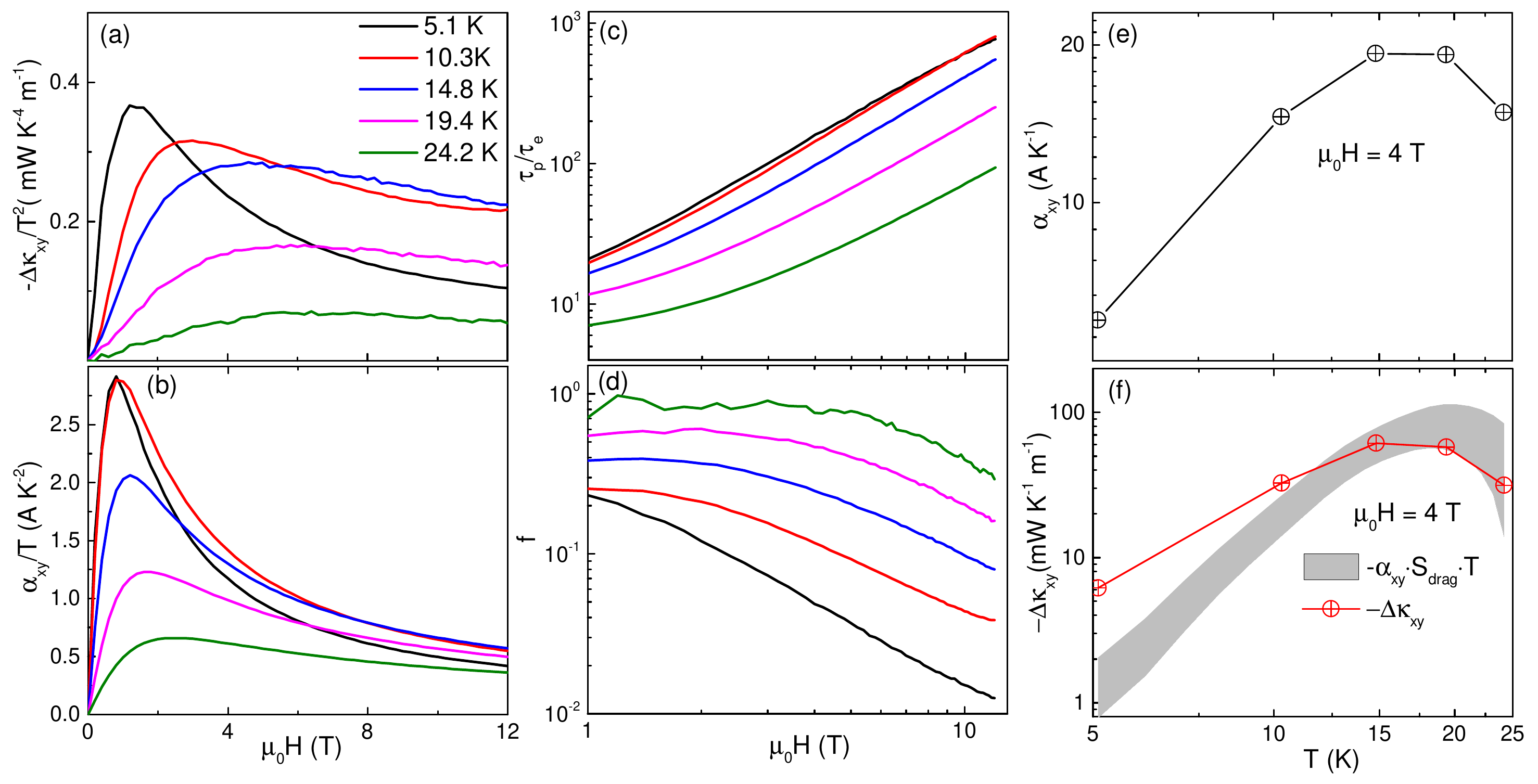}
\caption{\textbf{ Quantitative analysis of $\Delta \kappa_{xy}$:} (a) Field dependence of $\Delta\kappa_{xy}/T^3$ at different temperatures. (b) Field dependence of $\alpha_{xy}/T$ at different temperatures. In our picture, this is the main driver of the field dependence of $\Delta\kappa_{xy}$ (c) Field dependence of the ratio of the phonon to electron scattering time, extracted from electric and thermal conductivity data. Note that $\tau_P \gg \tau_e$ and the ratio enhances with magnetic field. (d) The field dependence of $f$, obtained by equating  $\Delta \kappa_{xy}$ and $\kappa_{xy}(drag)$ given by Eq. \ref{kappa_ep}.  (e), The temperature dependence of $\alpha_{xy}$ at 4 T. (f) The temperature dependence of   $-\Delta \kappa_{xy}$  and  $-\alpha_{xy} S_{drag}\cdot T$. The width of the latter represents the uncertainty in separating the phonon-drag and the diffusive components of the Seebeck coefficient (see the supplement \cite{SM}).}
\label{fig: fit}
\end{figure*}
 
\subsection*{Quantitative account of the data}
We proceed now to compare $\Delta \kappa_{xy}$ resolved by experiment, with $\kappa_{xy}(drag)$, expected by Eq. \ref{kappa_ep}. 
With the exception of $f$, all terms of  Eq. \ref{kappa_ep} are experimentally accessible. The sound velocity is $v_s=7.8 km/s$ \cite{REHWALD}, the effective mass of electrons is $m^\star= 1.8m_e$ \cite{Lin2013}. $\tau_p$  can be extracted from phonon thermal conductivity and $\tau_e$ from the electrical conductivity.

Fig. \ref{fig: fit}a shows  the  non-monotonous field dependence of $\kappa_{xy}(drag)$, which mirrors the field dependence of $\alpha_{xy}$ (Fig. \ref{fig: fit}b), which, after an initial increase, steadily decreases in the high-field regime. Since $S_{xx}> S_{xy}$ and $\sigma_{xy} \geq \sigma_{xx}$,  $\alpha_{xy}\simeq S_{xx}\sigma_{xy}$ and the field dependence of $\alpha_{xy}$, is similar to the field dependence of Hall conductivity, which steadily decreases in the high-field regime ($\sigma_{xy} (B\longrightarrow \infty) \longrightarrow 0$ ). 

The ratio of $\tau_p/\tau_e$, shown on Fig. \ref{fig: fit}c, is much larger than unity and steadily increases with magnetic field. This is because, as previously documented \cite{Collignon2021}, the mobility of electrons diminishes with magnetic field, presumably because partially extended disorder becomes more effective in scattering electrons with increasing magnetic field which confines the electron wave function. The phonon scattering time on the other hand is barely affected by magnetic field.

By assuming equality between $\Delta \kappa_{xy}$ and  $\kappa_{xy}(drag)$ and using Eq. \ref{kappa_ep}, we can extract $f$. The results are shown in Fig. \ref{fig: fit}d. We find that $f <1$, attesting the soundness of our approach. The efficiency of momentum transfer between phonons and electrons is close to 1 at the peak temperature and steadily decreases with temperature. As the temperature decreases, the relative frequency of electron-phonon scattering events decreases. We also note that a strong coupling between electrons and soft ferroelectric phonons has been invoked \cite{Kumar2021, Nazaryan2021} to explain the transport properties of metallic STO \cite{Lin2015,Collignon2020}. The steady field-induced decrease in $f$ indicates that the efficiency of the momentum transfer between  phonons and electrons decreases with the decrease in the electronic scattering time and the growing mismatch between the phonon and the electron scattering time.

 The soundness of our diagnostic can be checked by comparing  the extra thermal Hall effect and the phonon drag Seebeck effect, supposed to share the same origin. Both effects, measured by distinct experiments, peak around 20 K. Their amplitude match too.  Eq.\ref{Herring} and Eq.\ref{kappa_ep} together with the Kelvin relation ($\Pi= ST$) and the identification of $\Delta \kappa_{xy}$ with $\kappa_{xy}(drag)$ implies an equality between the two extracted quantities: $\Delta \kappa_{xy} \approx -\alpha_{xy} S_{drag} T$.  Fig. \ref{fig: fit} f compares $\Delta \kappa_{xy}$ and $-\alpha_{xy}\cdot S_{drag}\cdot T$ as a function of temperature at 4 T. The agreement between 25 K down to 10 K confirms the quantitative self-consistency of the data. The disagreement at 5 K indicates the limits of our approximations when $f$ strongly varies with magnetic field. 

\subsection*{Relevance to other metallic solids}
Our result implies that the combination of phonon-drag (frequently encountered in semiconductors) and a sizeable transverse thermoelectric conductivity, $\alpha_{xy}$ (known to become large when the carrier density is low  density and carrier mobility is high\cite{Behnia_2016}) can give rise to a phonon-drag THE. These conditions can be formulated in term of a hierarchy of time scales. The scattering time of phonons should exceed the scattering time of electrons and the latter should in turn exceed the inverse of the cyclotron frequency, in order to ensure longitudinal-to-transverse conversion. Thus, a sizeable phonon drag THE requires: 

\begin{equation}
\tau_p > \tau_e > \omega_c^{-1}
\label{cond1}
\end{equation}

The second condition is that the rate of momentum lost by phonons is to be comparable to the rate of phonon-electron exchange : 
\begin{equation}
  f \gg 0
  \label{cond2}
\end{equation}
Note that the latter condition can only be satisfied at finite temperature. The hydrodynamic window, identified by Gurzhi \cite{gurzhi1968} requires  a hierarchy of time scales too: The momentum-conserving scattering time should outweigh the boundary scattering time, itself larger than the momentum-relaxing scattering time. As a result, hydrodynamic features are expected in a limited temperature window and in a limited number of materials.

Inequalities \ref{cond1} and \ref{cond2} specify the conditions for expecting a phonon drag THE. Long $\tau_p$ is expected in a crystal at low temperature when phonon wavelength is long and point defects cannot scatter them.  In many dilute metals, when the Bohr radius of the semiconducting parent is long \cite{Behnia_2015} carriers are more mobile than in metallic silicon. As a consequence, the inequality $\tau_e > \omega_c^{-1}$ is easily satisfied at low fields. If electron-phonon momentum exchange happens to be frequent too, then  an effect similar to the one observed here is expected. Possible candidates are Bi$_2$Se$_3$ \cite{Fauque2013}, InAs \cite{Jaoui2020}, PbTe \cite{bookPbTe} and their sister compounds. It is not surprising that the thermal Hall effect in metallic cuprates is not detetectably amplified by phonon drag \cite{Grissonnanche2019}. In their case the mobility of carriers in  cuprates is low and, as a consequence, condition \ref{cond1} is not satisfied. Indeed, the  measured Hall angle in La$_{1-x}$Sr$_x$CuO$_4$ is quite small. At B=10 T, $\sigma_{xy}/\sigma_{xx}\ll 0.1$ \cite{Ando2004} and therefore $\omega_c\tau_e <1$. 

The result reported here does not provide a direct solution to the puzzle of phonon thermal Hall effect in insulators. Nevertheless, the complicity between two types of heat carriers, one with a long scattering time (here phonons) and another with a large Hall angle (here electrons) may have relevance to other contexts. One can imagine a scenario close to ours in an insulating solid hosting two types of carriers:  ordinary  phonons with a long scattering time and magnons or chiral phonons or with a short scattering time and a large transverse response.   
\bibliography{ref}

\begin{thebibliography}{56}%
\makeatletter
\providecommand \@ifxundefined [1]{%
 \@ifx{#1\undefined}
}%
\providecommand \@ifnum [1]{%
 \ifnum #1\expandafter \@firstoftwo
 \else \expandafter \@secondoftwo
 \fi
}%
\providecommand \@ifx [1]{%
 \ifx #1\expandafter \@firstoftwo
 \else \expandafter \@secondoftwo
 \fi
}%
\providecommand \natexlab [1]{#1}%
\providecommand \enquote  [1]{``#1''}%
\providecommand \bibnamefont  [1]{#1}%
\providecommand \bibfnamefont [1]{#1}%
\providecommand \citenamefont [1]{#1}%
\providecommand \href@noop [0]{\@secondoftwo}%
\providecommand \href [0]{\begingroup \@sanitize@url \@href}%
\providecommand \@href[1]{\@@startlink{#1}\@@href}%
\providecommand \@@href[1]{\endgroup#1\@@endlink}%
\providecommand \@sanitize@url [0]{\catcode `\\12\catcode `\$12\catcode
  `\&12\catcode `\#12\catcode `\^12\catcode `\_12\catcode `\%12\relax}%
\providecommand \@@startlink[1]{}%
\providecommand \@@endlink[0]{}%
\providecommand \url  [0]{\begingroup\@sanitize@url \@url }%
\providecommand \@url [1]{\endgroup\@href {#1}{\urlprefix }}%
\providecommand \urlprefix  [0]{URL }%
\providecommand \Eprint [0]{\href }%
\providecommand \doibase [0]{https://doi.org/}%
\providecommand \selectlanguage [0]{\@gobble}%
\providecommand \bibinfo  [0]{\@secondoftwo}%
\providecommand \bibfield  [0]{\@secondoftwo}%
\providecommand \translation [1]{[#1]}%
\providecommand \BibitemOpen [0]{}%
\providecommand \bibitemStop [0]{}%
\providecommand \bibitemNoStop [0]{.\EOS\space}%
\providecommand \EOS [0]{\spacefactor3000\relax}%
\providecommand \BibitemShut  [1]{\csname bibitem#1\endcsname}%
\let\auto@bib@innerbib\@empty
\bibitem [{\citenamefont {Sugii}\ \emph {et~al.}(2017)\citenamefont {Sugii},
  \citenamefont {Shimozawa}, \citenamefont {Watanabe}, \citenamefont {Suzuki},
  \citenamefont {Halim}, \citenamefont {Kimata}, \citenamefont {Matsumoto},
  \citenamefont {Nakatsuji},\ and\ \citenamefont {Yamashita}}]{Sugii2017}%
  \BibitemOpen
  \bibfield  {author} {\bibinfo {author} {\bibfnamefont {K.}~\bibnamefont
  {Sugii}}, \bibinfo {author} {\bibfnamefont {M.}~\bibnamefont {Shimozawa}},
  \bibinfo {author} {\bibfnamefont {D.}~\bibnamefont {Watanabe}}, \bibinfo
  {author} {\bibfnamefont {Y.}~\bibnamefont {Suzuki}}, \bibinfo {author}
  {\bibfnamefont {M.}~\bibnamefont {Halim}}, \bibinfo {author} {\bibfnamefont
  {M.}~\bibnamefont {Kimata}}, \bibinfo {author} {\bibfnamefont
  {Y.}~\bibnamefont {Matsumoto}}, \bibinfo {author} {\bibfnamefont
  {S.}~\bibnamefont {Nakatsuji}},\ and\ \bibinfo {author} {\bibfnamefont
  {M.}~\bibnamefont {Yamashita}},\ }\bibfield  {title} {\bibinfo {title}
  {Thermal {Hall} effect in a phonon-glass
  {${\mathrm{Ba}}_{3}{\mathrm{CuSb}}_{2}{\mathrm{O}}_{9}$}},\ }\href
  {https://doi.org/10.1103/PhysRevLett.118.145902} {\bibfield  {journal}
  {\bibinfo  {journal} {Phys. Rev. Lett.}\ }\textbf {\bibinfo {volume} {118}},\
  \bibinfo {pages} {145902} (\bibinfo {year} {2017})}\BibitemShut {NoStop}%
\bibitem [{\citenamefont {Kasahara}\ \emph {et~al.}(2018)\citenamefont
  {Kasahara}, \citenamefont {Sugii}, \citenamefont {Ohnishi}, \citenamefont
  {Shimozawa}, \citenamefont {Yamashita}, \citenamefont {Kurita}, \citenamefont
  {Tanaka}, \citenamefont {Nasu}, \citenamefont {Motome}, \citenamefont
  {Shibauchi},\ and\ \citenamefont {Matsuda}}]{Kasahara2018}%
  \BibitemOpen
  \bibfield  {author} {\bibinfo {author} {\bibfnamefont {Y.}~\bibnamefont
  {Kasahara}}, \bibinfo {author} {\bibfnamefont {K.}~\bibnamefont {Sugii}},
  \bibinfo {author} {\bibfnamefont {T.}~\bibnamefont {Ohnishi}}, \bibinfo
  {author} {\bibfnamefont {M.}~\bibnamefont {Shimozawa}}, \bibinfo {author}
  {\bibfnamefont {M.}~\bibnamefont {Yamashita}}, \bibinfo {author}
  {\bibfnamefont {N.}~\bibnamefont {Kurita}}, \bibinfo {author} {\bibfnamefont
  {H.}~\bibnamefont {Tanaka}}, \bibinfo {author} {\bibfnamefont
  {J.}~\bibnamefont {Nasu}}, \bibinfo {author} {\bibfnamefont {Y.}~\bibnamefont
  {Motome}}, \bibinfo {author} {\bibfnamefont {T.}~\bibnamefont {Shibauchi}},\
  and\ \bibinfo {author} {\bibfnamefont {Y.}~\bibnamefont {Matsuda}},\
  }\bibfield  {title} {\bibinfo {title} {Unusual thermal {{Hall}} effect in a
  kitaev spin liquid candidate
  {$\ensuremath{\alpha}\text{\ensuremath{-}}{\mathrm{RuCl}}_{3}$}},\ }\href
  {https://doi.org/10.1103/PhysRevLett.120.217205} {\bibfield  {journal}
  {\bibinfo  {journal} {Phys. Rev. Lett.}\ }\textbf {\bibinfo {volume} {120}},\
  \bibinfo {pages} {217205} (\bibinfo {year} {2018})}\BibitemShut {NoStop}%
\bibitem [{\citenamefont {Grissonnanche}\ \emph {et~al.}(2019)\citenamefont
  {Grissonnanche}, \citenamefont {Legros}, \citenamefont {Badoux},
  \citenamefont {Lefran{\c{c}}ois}, \citenamefont {Zatko}, \citenamefont
  {Lizaire}, \citenamefont {Lalibert{\'{e}}}, \citenamefont {Gourgout},
  \citenamefont {Zhou}, \citenamefont {Pyon}, \citenamefont {Takayama},
  \citenamefont {Takagi}, \citenamefont {Ono}, \citenamefont {Doiron-Leyraud},\
  and\ \citenamefont {Taillefer}}]{Grissonnanche2019}%
  \BibitemOpen
  \bibfield  {author} {\bibinfo {author} {\bibfnamefont {G.}~\bibnamefont
  {Grissonnanche}}, \bibinfo {author} {\bibfnamefont {A.}~\bibnamefont
  {Legros}}, \bibinfo {author} {\bibfnamefont {S.}~\bibnamefont {Badoux}},
  \bibinfo {author} {\bibfnamefont {E.}~\bibnamefont {Lefran{\c{c}}ois}},
  \bibinfo {author} {\bibfnamefont {V.}~\bibnamefont {Zatko}}, \bibinfo
  {author} {\bibfnamefont {M.}~\bibnamefont {Lizaire}}, \bibinfo {author}
  {\bibfnamefont {F.}~\bibnamefont {Lalibert{\'{e}}}}, \bibinfo {author}
  {\bibfnamefont {A.}~\bibnamefont {Gourgout}}, \bibinfo {author}
  {\bibfnamefont {J.-S.}\ \bibnamefont {Zhou}}, \bibinfo {author}
  {\bibfnamefont {S.}~\bibnamefont {Pyon}}, \bibinfo {author} {\bibfnamefont
  {T.}~\bibnamefont {Takayama}}, \bibinfo {author} {\bibfnamefont
  {H.}~\bibnamefont {Takagi}}, \bibinfo {author} {\bibfnamefont
  {S.}~\bibnamefont {Ono}}, \bibinfo {author} {\bibfnamefont {N.}~\bibnamefont
  {Doiron-Leyraud}},\ and\ \bibinfo {author} {\bibfnamefont {L.}~\bibnamefont
  {Taillefer}},\ }\bibfield  {title} {\bibinfo {title} {{Giant thermal {Hall}
  conductivity in the pseudogap phase of cuprate superconductors}},\ }\href
  {https://doi.org/10.1038/s41586-019-1375-0} {\bibfield  {journal} {\bibinfo
  {journal} {Nature}\ }\textbf {\bibinfo {volume} {571}},\ \bibinfo {pages}
  {376} (\bibinfo {year} {2019})}\BibitemShut {NoStop}%
\bibitem [{\citenamefont {Hirokane}\ \emph {et~al.}(2019)\citenamefont
  {Hirokane}, \citenamefont {Nii}, \citenamefont {Tomioka},\ and\ \citenamefont
  {Onose}}]{Hirokane2019}%
  \BibitemOpen
  \bibfield  {author} {\bibinfo {author} {\bibfnamefont {Y.}~\bibnamefont
  {Hirokane}}, \bibinfo {author} {\bibfnamefont {Y.}~\bibnamefont {Nii}},
  \bibinfo {author} {\bibfnamefont {Y.}~\bibnamefont {Tomioka}},\ and\ \bibinfo
  {author} {\bibfnamefont {Y.}~\bibnamefont {Onose}},\ }\bibfield  {title}
  {\bibinfo {title} {Phononic thermal {Hall} effect in diluted terbium
  oxides},\ }\href {https://doi.org/10.1103/PhysRevB.99.134419} {\bibfield
  {journal} {\bibinfo  {journal} {Phys. Rev. B}\ }\textbf {\bibinfo {volume}
  {99}},\ \bibinfo {pages} {134419} (\bibinfo {year} {2019})}\BibitemShut
  {NoStop}%
\bibitem [{\citenamefont {Li}\ \emph {et~al.}(2020)\citenamefont {Li},
  \citenamefont {Fauqu\'e}, \citenamefont {Zhu},\ and\ \citenamefont
  {Behnia}}]{Li2020}%
  \BibitemOpen
  \bibfield  {author} {\bibinfo {author} {\bibfnamefont {X.}~\bibnamefont
  {Li}}, \bibinfo {author} {\bibfnamefont {B.}~\bibnamefont {Fauqu\'e}},
  \bibinfo {author} {\bibfnamefont {Z.}~\bibnamefont {Zhu}},\ and\ \bibinfo
  {author} {\bibfnamefont {K.}~\bibnamefont {Behnia}},\ }\bibfield  {title}
  {\bibinfo {title} {Phonon thermal {{Hall}} effect in strontium titanate},\
  }\href {https://doi.org/10.1103/PhysRevLett.124.105901} {\bibfield  {journal}
  {\bibinfo  {journal} {Phys. Rev. Lett.}\ }\textbf {\bibinfo {volume} {124}},\
  \bibinfo {pages} {105901} (\bibinfo {year} {2020})}\BibitemShut {NoStop}%
\bibitem [{\citenamefont {Bruin}\ \emph {et~al.}(2021)\citenamefont {Bruin},
  \citenamefont {Claus}, \citenamefont {Matsumoto}, \citenamefont {Kurita},
  \citenamefont {Tanaka},\ and\ \citenamefont {Takagi}}]{bruin2021}%
  \BibitemOpen
  \bibfield  {author} {\bibinfo {author} {\bibfnamefont {J.~A.~N.}\
  \bibnamefont {Bruin}}, \bibinfo {author} {\bibfnamefont {R.~R.}\ \bibnamefont
  {Claus}}, \bibinfo {author} {\bibfnamefont {Y.}~\bibnamefont {Matsumoto}},
  \bibinfo {author} {\bibfnamefont {N.}~\bibnamefont {Kurita}}, \bibinfo
  {author} {\bibfnamefont {H.}~\bibnamefont {Tanaka}},\ and\ \bibinfo {author}
  {\bibfnamefont {H.}~\bibnamefont {Takagi}},\ }\href@noop {} {\bibinfo {title}
  {Robustness of the thermal {Hall} effect close to half-quantization in a
  field-induced spin liquid state}} (\bibinfo {year} {2021}),\ \Eprint
  {https://arxiv.org/abs/2104.12184} {arXiv:2104.12184 [cond-mat.str-el]}
  \BibitemShut {NoStop}%
\bibitem [{\citenamefont {{Lefran{\c{c}}ois}}\ \emph
  {et~al.}(2021)\citenamefont {{Lefran{\c{c}}ois}}, \citenamefont
  {{Grissonnanche}}, \citenamefont {{Baglo}}, \citenamefont {{Lampen-Kelley}},
  \citenamefont {{Yan}}, \citenamefont {{Balz}}, \citenamefont {{Mandrus}},
  \citenamefont {{Nagler}}, \citenamefont {{Kim}}, \citenamefont {{Kim}},
  \citenamefont {{Doiron-Leyraud}},\ and\ \citenamefont
  {{Taillefer}}}]{lefrancois2021}%
  \BibitemOpen
  \bibfield  {author} {\bibinfo {author} {\bibfnamefont {{\'E}.}~\bibnamefont
  {{Lefran{\c{c}}ois}}}, \bibinfo {author} {\bibfnamefont {G.}~\bibnamefont
  {{Grissonnanche}}}, \bibinfo {author} {\bibfnamefont {J.}~\bibnamefont
  {{Baglo}}}, \bibinfo {author} {\bibfnamefont {P.}~\bibnamefont
  {{Lampen-Kelley}}}, \bibinfo {author} {\bibfnamefont {J.}~\bibnamefont
  {{Yan}}}, \bibinfo {author} {\bibfnamefont {C.}~\bibnamefont {{Balz}}},
  \bibinfo {author} {\bibfnamefont {D.}~\bibnamefont {{Mandrus}}}, \bibinfo
  {author} {\bibfnamefont {S.~E.}\ \bibnamefont {{Nagler}}}, \bibinfo {author}
  {\bibfnamefont {S.}~\bibnamefont {{Kim}}}, \bibinfo {author} {\bibfnamefont
  {Y.-J.}\ \bibnamefont {{Kim}}}, \bibinfo {author} {\bibfnamefont
  {N.}~\bibnamefont {{Doiron-Leyraud}}},\ and\ \bibinfo {author} {\bibfnamefont
  {L.}~\bibnamefont {{Taillefer}}},\ }\bibfield  {title} {\bibinfo {title}
  {{Evidence of a Phonon {Hall} Effect in the Kitaev Spin Liquid Candidate
  {$\alpha$-RuCl$_3$}}},\ }\href@noop {} {\bibfield  {journal} {\bibinfo
  {journal} {arXiv e-prints}\ ,\ \bibinfo {eid} {arXiv:2111.05493}} (\bibinfo
  {year} {2021})},\ \Eprint {https://arxiv.org/abs/2111.05493}
  {arXiv:2111.05493 [cond-mat.str-el]} \BibitemShut {NoStop}%
\bibitem [{\citenamefont {Collignon}\ \emph {et~al.}(2019)\citenamefont
  {Collignon}, \citenamefont {Lin}, \citenamefont {Rischau}, \citenamefont
  {Fauqu{\'e}},\ and\ \citenamefont {Behnia}}]{Collignon2019}%
  \BibitemOpen
  \bibfield  {author} {\bibinfo {author} {\bibfnamefont {C.}~\bibnamefont
  {Collignon}}, \bibinfo {author} {\bibfnamefont {X.}~\bibnamefont {Lin}},
  \bibinfo {author} {\bibfnamefont {C.~W.}\ \bibnamefont {Rischau}}, \bibinfo
  {author} {\bibfnamefont {B.}~\bibnamefont {Fauqu{\'e}}},\ and\ \bibinfo
  {author} {\bibfnamefont {K.}~\bibnamefont {Behnia}},\ }\bibfield  {title}
  {\bibinfo {title} {Metallicity and superconductivity in doped strontium
  titanate},\ }\href@noop {} {\bibfield  {journal} {\bibinfo  {journal} {Annual
  Review of Condensed Matter Physics}\ }\textbf {\bibinfo {volume} {10}},\
  \bibinfo {pages} {25} (\bibinfo {year} {2019})}\BibitemShut {NoStop}%
\bibitem [{\citenamefont {Sim}\ \emph {et~al.}(2021)\citenamefont {Sim},
  \citenamefont {Yang}, \citenamefont {Kim}, \citenamefont {Coak},
  \citenamefont {Itoh}, \citenamefont {Noda},\ and\ \citenamefont
  {Park}}]{Sim2021}%
  \BibitemOpen
  \bibfield  {author} {\bibinfo {author} {\bibfnamefont {S.}~\bibnamefont
  {Sim}}, \bibinfo {author} {\bibfnamefont {H.}~\bibnamefont {Yang}}, \bibinfo
  {author} {\bibfnamefont {H.-L.}\ \bibnamefont {Kim}}, \bibinfo {author}
  {\bibfnamefont {M.~J.}\ \bibnamefont {Coak}}, \bibinfo {author}
  {\bibfnamefont {M.}~\bibnamefont {Itoh}}, \bibinfo {author} {\bibfnamefont
  {Y.}~\bibnamefont {Noda}},\ and\ \bibinfo {author} {\bibfnamefont {J.-G.}\
  \bibnamefont {Park}},\ }\bibfield  {title} {\bibinfo {title} {Sizable
  suppression of thermal {Hall} effect upon isotopic substitution in
  $\mathrm{SrTiO_{3}}$},\ }\href
  {https://doi.org/10.1103/PhysRevLett.126.015901} {\bibfield  {journal}
  {\bibinfo  {journal} {Phys. Rev. Lett.}\ }\textbf {\bibinfo {volume} {126}},\
  \bibinfo {pages} {015901} (\bibinfo {year} {2021})}\BibitemShut {NoStop}%
\bibitem [{\citenamefont {Qin}\ \emph {et~al.}(2012)\citenamefont {Qin},
  \citenamefont {Zhou},\ and\ \citenamefont {Shi}}]{Shin2012}%
  \BibitemOpen
  \bibfield  {author} {\bibinfo {author} {\bibfnamefont {T.}~\bibnamefont
  {Qin}}, \bibinfo {author} {\bibfnamefont {J.}~\bibnamefont {Zhou}},\ and\
  \bibinfo {author} {\bibfnamefont {J.}~\bibnamefont {Shi}},\ }\bibfield
  {title} {\bibinfo {title} {Berry curvature and the phonon {Hall} effect},\
  }\href {https://doi.org/10.1103/PhysRevB.86.104305} {\bibfield  {journal}
  {\bibinfo  {journal} {Phys. Rev. B}\ }\textbf {\bibinfo {volume} {86}},\
  \bibinfo {pages} {104305} (\bibinfo {year} {2012})}\BibitemShut {NoStop}%
\bibitem [{\citenamefont {Chen}\ \emph {et~al.}(2020)\citenamefont {Chen},
  \citenamefont {Kivelson},\ and\ \citenamefont {Sun}}]{Chen2020}%
  \BibitemOpen
  \bibfield  {author} {\bibinfo {author} {\bibfnamefont {J.-Y.}\ \bibnamefont
  {Chen}}, \bibinfo {author} {\bibfnamefont {S.~A.}\ \bibnamefont {Kivelson}},\
  and\ \bibinfo {author} {\bibfnamefont {X.-Q.}\ \bibnamefont {Sun}},\
  }\bibfield  {title} {\bibinfo {title} {Enhanced thermal {Hall} effect in
  nearly ferroelectric insulators},\ }\href
  {https://doi.org/10.1103/PhysRevLett.124.167601} {\bibfield  {journal}
  {\bibinfo  {journal} {Phys. Rev. Lett.}\ }\textbf {\bibinfo {volume} {124}},\
  \bibinfo {pages} {167601} (\bibinfo {year} {2020})}\BibitemShut {NoStop}%
\bibitem [{\citenamefont {{Sun}}\ \emph {et~al.}(2021)\citenamefont {{Sun}},
  \citenamefont {{Chen}},\ and\ \citenamefont {{Kivelson}}}]{Sun2021b}%
  \BibitemOpen
  \bibfield  {author} {\bibinfo {author} {\bibfnamefont {X.-Q.}\ \bibnamefont
  {{Sun}}}, \bibinfo {author} {\bibfnamefont {J.-Y.}\ \bibnamefont {{Chen}}},\
  and\ \bibinfo {author} {\bibfnamefont {S.~A.}\ \bibnamefont {{Kivelson}}},\
  }\bibfield  {title} {\bibinfo {title} {{Large extrinsic phonon thermal {Hall}
  effect from resonant scattering}},\ }\href@noop {} {\bibfield  {journal}
  {\bibinfo  {journal} {arXiv e-prints}\ ,\ \bibinfo {eid} {arXiv:2109.12117}}
  (\bibinfo {year} {2021})},\ \Eprint {https://arxiv.org/abs/2109.12117}
  {arXiv:2109.12117 [cond-mat.mes-hall]} \BibitemShut {NoStop}%
\bibitem [{\citenamefont {{Flebus}}\ and\ \citenamefont
  {{MacDonald}}(2021)}]{Flebus2021}%
  \BibitemOpen
  \bibfield  {author} {\bibinfo {author} {\bibfnamefont {B.}~\bibnamefont
  {{Flebus}}}\ and\ \bibinfo {author} {\bibfnamefont {A.~H.}\ \bibnamefont
  {{MacDonald}}},\ }\bibfield  {title} {\bibinfo {title} {{Charged Defects and
  Phonon {Hall} Effects in Ionic Crystals}},\ }\href@noop {} {\bibfield
  {journal} {\bibinfo  {journal} {arXiv e-prints}\ ,\ \bibinfo {eid}
  {arXiv:2106.13889}} (\bibinfo {year} {2021})},\ \Eprint
  {https://arxiv.org/abs/2106.13889} {arXiv:2106.13889 [cond-mat.mes-hall]}
  \BibitemShut {NoStop}%
\bibitem [{\citenamefont {Guo}\ and\ \citenamefont {Sachdev}(2021)}]{Guo2021}%
  \BibitemOpen
  \bibfield  {author} {\bibinfo {author} {\bibfnamefont {H.}~\bibnamefont
  {Guo}}\ and\ \bibinfo {author} {\bibfnamefont {S.}~\bibnamefont {Sachdev}},\
  }\bibfield  {title} {\bibinfo {title} {Extrinsic phonon thermal {Hall}
  transport from {Hall} viscosity},\ }\href
  {https://doi.org/10.1103/PhysRevB.103.205115} {\bibfield  {journal} {\bibinfo
   {journal} {Phys. Rev. B}\ }\textbf {\bibinfo {volume} {103}},\ \bibinfo
  {pages} {205115} (\bibinfo {year} {2021})}\BibitemShut {NoStop}%
\bibitem [{\citenamefont {{Zabalo}}\ \emph {et~al.}(2021)\citenamefont
  {{Zabalo}}, \citenamefont {{Dreyer}},\ and\ \citenamefont
  {{Stengel}}}]{Zabalo2021}%
  \BibitemOpen
  \bibfield  {author} {\bibinfo {author} {\bibfnamefont {A.}~\bibnamefont
  {{Zabalo}}}, \bibinfo {author} {\bibfnamefont {C.~E.}\ \bibnamefont
  {{Dreyer}}},\ and\ \bibinfo {author} {\bibfnamefont {M.}~\bibnamefont
  {{Stengel}}},\ }\bibfield  {title} {\bibinfo {title} {{Rotational $g$ factors
  and Lorentz forces of molecules and solids from density-functional
  perturbation theory}},\ }\href@noop {} {\bibfield  {journal} {\bibinfo
  {journal} {arXiv e-prints}\ ,\ \bibinfo {eid} {arXiv:2112.11946}} (\bibinfo
  {year} {2021})},\ \Eprint {https://arxiv.org/abs/2112.11946}
  {arXiv:2112.11946 [cond-mat.mtrl-sci]} \BibitemShut {NoStop}%
\bibitem [{\citenamefont {Bhalla}\ and\ \citenamefont
  {Das}(2021)}]{Bhalla_2021}%
  \BibitemOpen
  \bibfield  {author} {\bibinfo {author} {\bibfnamefont {P.}~\bibnamefont
  {Bhalla}}\ and\ \bibinfo {author} {\bibfnamefont {N.}~\bibnamefont {Das}},\
  }\bibfield  {title} {\bibinfo {title} {Optical phonon contribution to the
  thermal conductivity of a quantum paraelectric},\ }\href
  {https://doi.org/10.1088/1361-648x/ac08b7} {\bibfield  {journal} {\bibinfo
  {journal} {Journal of Physics: Condensed Matter}\ }\textbf {\bibinfo {volume}
  {33}},\ \bibinfo {pages} {345401} (\bibinfo {year} {2021})}\BibitemShut
  {NoStop}%
\bibitem [{\citenamefont {Rowley}\ \emph {et~al.}(2014)\citenamefont {Rowley},
  \citenamefont {Spalek}, \citenamefont {Smith}, \citenamefont {Dean},
  \citenamefont {Itoh}, \citenamefont {Scott}, \citenamefont {Lonzarich},\ and\
  \citenamefont {Saxena}}]{Rowley2014}%
  \BibitemOpen
  \bibfield  {author} {\bibinfo {author} {\bibfnamefont {S.~E.}\ \bibnamefont
  {Rowley}}, \bibinfo {author} {\bibfnamefont {L.~J.}\ \bibnamefont {Spalek}},
  \bibinfo {author} {\bibfnamefont {R.~P.}\ \bibnamefont {Smith}}, \bibinfo
  {author} {\bibfnamefont {M.~P.~M.}\ \bibnamefont {Dean}}, \bibinfo {author}
  {\bibfnamefont {M.}~\bibnamefont {Itoh}}, \bibinfo {author} {\bibfnamefont
  {J.~F.}\ \bibnamefont {Scott}}, \bibinfo {author} {\bibfnamefont {G.~G.}\
  \bibnamefont {Lonzarich}},\ and\ \bibinfo {author} {\bibfnamefont {S.~S.}\
  \bibnamefont {Saxena}},\ }\bibfield  {title} {\bibinfo {title} {Ferroelectric
  quantum criticality},\ }\href {https://doi.org/10.1038/nphys2924} {\bibfield
  {journal} {\bibinfo  {journal} {Nature Physics}\ }\textbf {\bibinfo {volume}
  {10}},\ \bibinfo {pages} {367} (\bibinfo {year} {2014})}\BibitemShut
  {NoStop}%
\bibitem [{\citenamefont {M\"uller}\ and\ \citenamefont
  {Burkard}(1979)}]{Muller1979}%
  \BibitemOpen
  \bibfield  {author} {\bibinfo {author} {\bibfnamefont {K.~A.}\ \bibnamefont
  {M\"uller}}\ and\ \bibinfo {author} {\bibfnamefont {H.}~\bibnamefont
  {Burkard}},\ }\bibfield  {title} {\bibinfo {title} {Srti${\mathrm{o}}_{3}$:
  An intrinsic quantum paraelectric below 4 k},\ }\href
  {https://doi.org/10.1103/PhysRevB.19.3593} {\bibfield  {journal} {\bibinfo
  {journal} {Phys. Rev. B}\ }\textbf {\bibinfo {volume} {19}},\ \bibinfo
  {pages} {3593} (\bibinfo {year} {1979})}\BibitemShut {NoStop}%
\bibitem [{\citenamefont {Bednorz}\ and\ \citenamefont
  {M\"uller}(1984)}]{Bednorz1984}%
  \BibitemOpen
  \bibfield  {author} {\bibinfo {author} {\bibfnamefont {J.~G.}\ \bibnamefont
  {Bednorz}}\ and\ \bibinfo {author} {\bibfnamefont {K.~A.}\ \bibnamefont
  {M\"uller}},\ }\bibfield  {title} {\bibinfo {title}
  {{${\mathrm{Sr}}_{1\ensuremath{-}x}{\mathrm{Ca}}_{x}\mathrm{Ti}{\mathrm{O}}_{3}$}:
  An $\mathrm{XY}$ quantum ferroelectric with transition to randomness},\
  }\href {https://doi.org/10.1103/PhysRevLett.52.2289} {\bibfield  {journal}
  {\bibinfo  {journal} {Phys. Rev. Lett.}\ }\textbf {\bibinfo {volume} {52}},\
  \bibinfo {pages} {2289} (\bibinfo {year} {1984})}\BibitemShut {NoStop}%
\bibitem [{\citenamefont {Lemanov}\ \emph {et~al.}(1996)\citenamefont
  {Lemanov}, \citenamefont {Smirnova}, \citenamefont {Syrnikov},\ and\
  \citenamefont {Tarakanov}}]{Lemanov1996}%
  \BibitemOpen
  \bibfield  {author} {\bibinfo {author} {\bibfnamefont {V.~V.}\ \bibnamefont
  {Lemanov}}, \bibinfo {author} {\bibfnamefont {E.~P.}\ \bibnamefont
  {Smirnova}}, \bibinfo {author} {\bibfnamefont {P.~P.}\ \bibnamefont
  {Syrnikov}},\ and\ \bibinfo {author} {\bibfnamefont {E.~A.}\ \bibnamefont
  {Tarakanov}},\ }\bibfield  {title} {\bibinfo {title} {Phase transitions and
  glasslike behavior in
  {${\mathrm{Sr}}_{1\mathrm{\ensuremath{-}}\mathit{x}}$${\mathrm{Ba}}_{\mathit{x}}$${\mathrm{TiO}}_{3}$}},\
  }\href {https://doi.org/10.1103/PhysRevB.54.3151} {\bibfield  {journal}
  {\bibinfo  {journal} {Phys. Rev. B}\ }\textbf {\bibinfo {volume} {54}},\
  \bibinfo {pages} {3151} (\bibinfo {year} {1996})}\BibitemShut {NoStop}%
\bibitem [{\citenamefont {Itoh}\ \emph {et~al.}(1999)\citenamefont {Itoh},
  \citenamefont {Wang}, \citenamefont {Inaguma}, \citenamefont {Yamaguchi},
  \citenamefont {Shan},\ and\ \citenamefont {Nakamura}}]{Itoh1999}%
  \BibitemOpen
  \bibfield  {author} {\bibinfo {author} {\bibfnamefont {M.}~\bibnamefont
  {Itoh}}, \bibinfo {author} {\bibfnamefont {R.}~\bibnamefont {Wang}}, \bibinfo
  {author} {\bibfnamefont {Y.}~\bibnamefont {Inaguma}}, \bibinfo {author}
  {\bibfnamefont {T.}~\bibnamefont {Yamaguchi}}, \bibinfo {author}
  {\bibfnamefont {Y.-J.}\ \bibnamefont {Shan}},\ and\ \bibinfo {author}
  {\bibfnamefont {T.}~\bibnamefont {Nakamura}},\ }\bibfield  {title} {\bibinfo
  {title} {Ferroelectricity induced by oxygen isotope exchange in strontium
  titanate perovskite},\ }\href {https://doi.org/10.1103/PhysRevLett.82.3540}
  {\bibfield  {journal} {\bibinfo  {journal} {Phys. Rev. Lett.}\ }\textbf
  {\bibinfo {volume} {82}},\ \bibinfo {pages} {3540} (\bibinfo {year}
  {1999})}\BibitemShut {NoStop}%
\bibitem [{\citenamefont {Carpenter}\ \emph {et~al.}(2006)\citenamefont
  {Carpenter}, \citenamefont {Howard}, \citenamefont {Knight},\ and\
  \citenamefont {Zhang}}]{Carpenter2006}%
  \BibitemOpen
  \bibfield  {author} {\bibinfo {author} {\bibfnamefont {M.~A.}\ \bibnamefont
  {Carpenter}}, \bibinfo {author} {\bibfnamefont {C.~J.}\ \bibnamefont
  {Howard}}, \bibinfo {author} {\bibfnamefont {K.}~\bibnamefont {Knight}},\
  and\ \bibinfo {author} {\bibfnamefont {Z.}~\bibnamefont {Zhang}},\ }\bibfield
   {title} {\bibinfo {title} {Structural relationships and a phase diagram for
  $\mathrm{Sr_{1-x}Ca_{x}TiO_{3}}$ perovskites},\ }\href
  {https://doi.org/10.1088/0953-8984/18/48/002} {\bibfield  {journal} {\bibinfo
   {journal} {Journal of Physics: Condensed Matter}\ }\textbf {\bibinfo
  {volume} {18}},\ \bibinfo {pages} {10725} (\bibinfo {year}
  {2006})}\BibitemShut {NoStop}%
\bibitem [{\citenamefont {Lin}\ \emph {et~al.}(2013)\citenamefont {Lin},
  \citenamefont {Zhu}, \citenamefont {Fauqu\'e},\ and\ \citenamefont
  {Behnia}}]{Lin2013}%
  \BibitemOpen
  \bibfield  {author} {\bibinfo {author} {\bibfnamefont {X.}~\bibnamefont
  {Lin}}, \bibinfo {author} {\bibfnamefont {Z.}~\bibnamefont {Zhu}}, \bibinfo
  {author} {\bibfnamefont {B.}~\bibnamefont {Fauqu\'e}},\ and\ \bibinfo
  {author} {\bibfnamefont {K.}~\bibnamefont {Behnia}},\ }\bibfield  {title}
  {\bibinfo {title} {Fermi surface of the most dilute superconductor},\ }\href
  {https://doi.org/10.1103/PhysRevX.3.021002} {\bibfield  {journal} {\bibinfo
  {journal} {Phys. Rev. X}\ }\textbf {\bibinfo {volume} {3}},\ \bibinfo {pages}
  {021002} (\bibinfo {year} {2013})}\BibitemShut {NoStop}%
\bibitem [{\citenamefont {Wang}\ \emph {et~al.}(2019)\citenamefont {Wang},
  \citenamefont {Yang}, \citenamefont {Rischau}, \citenamefont {Xu},
  \citenamefont {Ren}, \citenamefont {Lorenz}, \citenamefont {Hemberger},
  \citenamefont {Lin},\ and\ \citenamefont {Behnia}}]{Wang2019}%
  \BibitemOpen
  \bibfield  {author} {\bibinfo {author} {\bibfnamefont {J.}~\bibnamefont
  {Wang}}, \bibinfo {author} {\bibfnamefont {L.}~\bibnamefont {Yang}}, \bibinfo
  {author} {\bibfnamefont {C.~W.}\ \bibnamefont {Rischau}}, \bibinfo {author}
  {\bibfnamefont {Z.}~\bibnamefont {Xu}}, \bibinfo {author} {\bibfnamefont
  {Z.}~\bibnamefont {Ren}}, \bibinfo {author} {\bibfnamefont {T.}~\bibnamefont
  {Lorenz}}, \bibinfo {author} {\bibfnamefont {J.}~\bibnamefont {Hemberger}},
  \bibinfo {author} {\bibfnamefont {X.}~\bibnamefont {Lin}},\ and\ \bibinfo
  {author} {\bibfnamefont {K.}~\bibnamefont {Behnia}},\ }\bibfield  {title}
  {\bibinfo {title} {Charge transport in a polar metal},\ }\href
  {https://doi.org/10.1038/s41535-019-0200-1} {\bibfield  {journal} {\bibinfo
  {journal} {npj Quantum Materials}\ }\textbf {\bibinfo {volume} {4}},\
  \bibinfo {pages} {61} (\bibinfo {year} {2019})}\BibitemShut {NoStop}%
\bibitem [{\citenamefont {Engelmayer}\ \emph {et~al.}(2019)\citenamefont
  {Engelmayer}, \citenamefont {Lin}, \citenamefont
  {Ko\ifmmode~\mbox{\c{c}}\else \c{c}\fi{}}, \citenamefont {Grams},
  \citenamefont {Hemberger}, \citenamefont {Behnia},\ and\ \citenamefont
  {Lorenz}}]{Engelmayer2019}%
  \BibitemOpen
  \bibfield  {author} {\bibinfo {author} {\bibfnamefont {J.}~\bibnamefont
  {Engelmayer}}, \bibinfo {author} {\bibfnamefont {X.}~\bibnamefont {Lin}},
  \bibinfo {author} {\bibfnamefont {F.}~\bibnamefont
  {Ko\ifmmode~\mbox{\c{c}}\else \c{c}\fi{}}}, \bibinfo {author} {\bibfnamefont
  {C.~P.}\ \bibnamefont {Grams}}, \bibinfo {author} {\bibfnamefont
  {J.}~\bibnamefont {Hemberger}}, \bibinfo {author} {\bibfnamefont
  {K.}~\bibnamefont {Behnia}},\ and\ \bibinfo {author} {\bibfnamefont
  {T.}~\bibnamefont {Lorenz}},\ }\bibfield  {title} {\bibinfo {title}
  {Ferroelectric order versus metallicity in
  $\mathrm{Sr_{1-x}Ca_{x}TiO_{3-\delta}}$ ($x=0.009$)},\ }\href
  {https://doi.org/10.1103/PhysRevB.100.195121} {\bibfield  {journal} {\bibinfo
   {journal} {Phys. Rev. B}\ }\textbf {\bibinfo {volume} {100}},\ \bibinfo
  {pages} {195121} (\bibinfo {year} {2019})}\BibitemShut {NoStop}%
\bibitem [{\citenamefont {Rischau}\ \emph {et~al.}(2017)\citenamefont
  {Rischau}, \citenamefont {Lin}, \citenamefont {Grams}, \citenamefont {Finck},
  \citenamefont {Harms}, \citenamefont {Engelmayer}, \citenamefont {Lorenz},
  \citenamefont {Gallais}, \citenamefont {Fauqu{\'e}}, \citenamefont
  {Hemberger},\ and\ \citenamefont {Behnia}}]{Rischau2017}%
  \BibitemOpen
  \bibfield  {author} {\bibinfo {author} {\bibfnamefont {C.~W.}\ \bibnamefont
  {Rischau}}, \bibinfo {author} {\bibfnamefont {X.}~\bibnamefont {Lin}},
  \bibinfo {author} {\bibfnamefont {C.~P.}\ \bibnamefont {Grams}}, \bibinfo
  {author} {\bibfnamefont {D.}~\bibnamefont {Finck}}, \bibinfo {author}
  {\bibfnamefont {S.}~\bibnamefont {Harms}}, \bibinfo {author} {\bibfnamefont
  {J.}~\bibnamefont {Engelmayer}}, \bibinfo {author} {\bibfnamefont
  {T.}~\bibnamefont {Lorenz}}, \bibinfo {author} {\bibfnamefont
  {Y.}~\bibnamefont {Gallais}}, \bibinfo {author} {\bibfnamefont
  {B.}~\bibnamefont {Fauqu{\'e}}}, \bibinfo {author} {\bibfnamefont
  {J.}~\bibnamefont {Hemberger}},\ and\ \bibinfo {author} {\bibfnamefont
  {K.}~\bibnamefont {Behnia}},\ }\bibfield  {title} {\bibinfo {title} {A
  ferroelectric quantum phase transition inside the superconducting dome of
  $\mathrm{Sr_{1-x}Ca_{x}TiO_{3-\delta}}$},\ }\href
  {https://doi.org/10.1038/nphys4085} {\bibfield  {journal} {\bibinfo
  {journal} {Nature Physics}\ }\textbf {\bibinfo {volume} {13}},\ \bibinfo
  {pages} {643} (\bibinfo {year} {2017})}\BibitemShut {NoStop}%
\bibitem [{\citenamefont {Herring}(1954)}]{Herring1954}%
  \BibitemOpen
  \bibfield  {author} {\bibinfo {author} {\bibfnamefont {C.}~\bibnamefont
  {Herring}},\ }\bibfield  {title} {\bibinfo {title} {Theory of the
  thermoelectric power of semiconductors},\ }\href
  {https://doi.org/10.1103/PhysRev.96.1163} {\bibfield  {journal} {\bibinfo
  {journal} {Phys. Rev.}\ }\textbf {\bibinfo {volume} {96}},\ \bibinfo {pages}
  {1163} (\bibinfo {year} {1954})}\BibitemShut {NoStop}%
\bibitem [{\citenamefont {Gurevich}\ and\ \citenamefont
  {Mashkevich}(1989)}]{Gurevich1989}%
  \BibitemOpen
  \bibfield  {author} {\bibinfo {author} {\bibfnamefont {Y.}~\bibnamefont
  {Gurevich}}\ and\ \bibinfo {author} {\bibfnamefont {O.}~\bibnamefont
  {Mashkevich}},\ }\bibfield  {title} {\bibinfo {title} {The electron-phonon
  drag and transport phenomena in semiconductors},\ }\href
  {https://doi.org/https://doi.org/10.1016/0370-1573(89)90011-2} {\bibfield
  {journal} {\bibinfo  {journal} {Physics Reports}\ }\textbf {\bibinfo {volume}
  {181}},\ \bibinfo {pages} {327} (\bibinfo {year} {1989})}\BibitemShut
  {NoStop}%
\bibitem [{\citenamefont {MacDonald}(2006)}]{MacDonald}%
  \BibitemOpen
  \bibfield  {author} {\bibinfo {author} {\bibfnamefont {D.~K.~C.}\
  \bibnamefont {MacDonald}},\ }\href@noop {} {\emph {\bibinfo {title}
  {Thermoelectricity : an introduction to the principles}}}\ (\bibinfo
  {publisher} {Dover Publications},\ \bibinfo {year} {2006})\BibitemShut
  {NoStop}%
\bibitem [{\citenamefont {Gurzhi}(1968)}]{gurzhi1968}%
  \BibitemOpen
  \bibfield  {author} {\bibinfo {author} {\bibfnamefont {R.~N.}\ \bibnamefont
  {Gurzhi}},\ }\bibfield  {title} {\bibinfo {title} {Hydrodynamic effects at
  low temperature},\ }\href {https://doi.org/10.1070/pu1968v011n02abeh003815}
  {\bibfield  {journal} {\bibinfo  {journal} {Soviet Physics Uspekhi}\ }\textbf
  {\bibinfo {volume} {11}},\ \bibinfo {pages} {255} (\bibinfo {year}
  {1968})}\BibitemShut {NoStop}%
\bibitem [{\citenamefont {Jaoui}\ \emph {et~al.}(2021)\citenamefont {Jaoui},
  \citenamefont {Gourgout}, \citenamefont {Seyfarth}, \citenamefont {Subedi},
  \citenamefont {Lorenz}, \citenamefont {Fauqué},\ and\ \citenamefont
  {Behnia}}]{jaoui2022}%
  \BibitemOpen
  \bibfield  {author} {\bibinfo {author} {\bibfnamefont {A.}~\bibnamefont
  {Jaoui}}, \bibinfo {author} {\bibfnamefont {A.}~\bibnamefont {Gourgout}},
  \bibinfo {author} {\bibfnamefont {G.}~\bibnamefont {Seyfarth}}, \bibinfo
  {author} {\bibfnamefont {A.}~\bibnamefont {Subedi}}, \bibinfo {author}
  {\bibfnamefont {T.}~\bibnamefont {Lorenz}}, \bibinfo {author} {\bibfnamefont
  {B.}~\bibnamefont {Fauqué}},\ and\ \bibinfo {author} {\bibfnamefont
  {K.}~\bibnamefont {Behnia}},\ }\bibfield  {title} {\bibinfo {title}
  {Formation of an electron-phonon bi-fluid in bulk antimony},\ }\href@noop {}
  {\bibfield  {journal} {\bibinfo  {journal} {arXiv}\ }\textbf {\bibinfo
  {volume} {2105.08408}} (\bibinfo {year} {2021})}\BibitemShut {NoStop}%
\bibitem [{\citenamefont {Levchenko}\ and\ \citenamefont
  {Schmalian}(2020)}]{levchenko2020}%
  \BibitemOpen
  \bibfield  {author} {\bibinfo {author} {\bibfnamefont {A.}~\bibnamefont
  {Levchenko}}\ and\ \bibinfo {author} {\bibfnamefont {J.}~\bibnamefont
  {Schmalian}},\ }\bibfield  {title} {\bibinfo {title} {Transport properties of
  strongly coupled electron–phonon liquids},\ }\href
  {https://doi.org/https://doi.org/10.1016/j.aop.2020.168218} {\bibfield
  {journal} {\bibinfo  {journal} {Annals of Physics}\ }\textbf {\bibinfo
  {volume} {419}},\ \bibinfo {pages} {168218} (\bibinfo {year}
  {2020})}\BibitemShut {NoStop}%
\bibitem [{\citenamefont {Huang}\ and\ \citenamefont
  {Lucas}(2021)}]{lucas2021}%
  \BibitemOpen
  \bibfield  {author} {\bibinfo {author} {\bibfnamefont {X.}~\bibnamefont
  {Huang}}\ and\ \bibinfo {author} {\bibfnamefont {A.}~\bibnamefont {Lucas}},\
  }\bibfield  {title} {\bibinfo {title} {Electron-phonon hydrodynamics},\
  }\href {https://doi.org/10.1103/PhysRevB.103.155128} {\bibfield  {journal}
  {\bibinfo  {journal} {Phys. Rev. B}\ }\textbf {\bibinfo {volume} {103}},\
  \bibinfo {pages} {155128} (\bibinfo {year} {2021})}\BibitemShut {NoStop}%
\bibitem [{\citenamefont {Callaway}(1959)}]{Callaway1959}%
  \BibitemOpen
  \bibfield  {author} {\bibinfo {author} {\bibfnamefont {J.}~\bibnamefont
  {Callaway}},\ }\bibfield  {title} {\bibinfo {title} {Model for lattice
  thermal conductivity at low temperatures},\ }\href
  {https://doi.org/10.1103/PhysRev.113.1046} {\bibfield  {journal} {\bibinfo
  {journal} {Phys. Rev.}\ }\textbf {\bibinfo {volume} {113}},\ \bibinfo {pages}
  {1046} (\bibinfo {year} {1959})}\BibitemShut {NoStop}%
\bibitem [{\citenamefont {Martelli}\ \emph {et~al.}(2018)\citenamefont
  {Martelli}, \citenamefont {Jim\'enez}, \citenamefont {Continentino},
  \citenamefont {Baggio-Saitovitch},\ and\ \citenamefont
  {Behnia}}]{Martlelli2018}%
  \BibitemOpen
  \bibfield  {author} {\bibinfo {author} {\bibfnamefont {V.}~\bibnamefont
  {Martelli}}, \bibinfo {author} {\bibfnamefont {J.~L.}\ \bibnamefont
  {Jim\'enez}}, \bibinfo {author} {\bibfnamefont {M.}~\bibnamefont
  {Continentino}}, \bibinfo {author} {\bibfnamefont {E.}~\bibnamefont
  {Baggio-Saitovitch}},\ and\ \bibinfo {author} {\bibfnamefont
  {K.}~\bibnamefont {Behnia}},\ }\bibfield  {title} {\bibinfo {title} {Thermal
  transport and phonon hydrodynamics in strontium titanate},\ }\href
  {https://doi.org/10.1103/PhysRevLett.120.125901} {\bibfield  {journal}
  {\bibinfo  {journal} {Phys. Rev. Lett.}\ }\textbf {\bibinfo {volume} {120}},\
  \bibinfo {pages} {125901} (\bibinfo {year} {2018})}\BibitemShut {NoStop}%
\bibitem [{\citenamefont {Rischau}\ \emph {et~al.}(2022)\citenamefont
  {Rischau}, \citenamefont {Pulmannov\'a}, \citenamefont {Scheerer},
  \citenamefont {Stucky}, \citenamefont {Giannini},\ and\ \citenamefont
  {van~der Marel}}]{Rischau2022}%
  \BibitemOpen
  \bibfield  {author} {\bibinfo {author} {\bibfnamefont {C.~W.}\ \bibnamefont
  {Rischau}}, \bibinfo {author} {\bibfnamefont {D.}~\bibnamefont
  {Pulmannov\'a}}, \bibinfo {author} {\bibfnamefont {G.~W.}\ \bibnamefont
  {Scheerer}}, \bibinfo {author} {\bibfnamefont {A.}~\bibnamefont {Stucky}},
  \bibinfo {author} {\bibfnamefont {E.}~\bibnamefont {Giannini}},\ and\
  \bibinfo {author} {\bibfnamefont {D.}~\bibnamefont {van~der Marel}},\
  }\bibfield  {title} {\bibinfo {title} {Isotope tuning of the superconducting
  dome of strontium titanate},\ }\href
  {https://doi.org/10.1103/PhysRevResearch.4.013019} {\bibfield  {journal}
  {\bibinfo  {journal} {Phys. Rev. Research}\ }\textbf {\bibinfo {volume}
  {4}},\ \bibinfo {pages} {013019} (\bibinfo {year} {2022})}\BibitemShut
  {NoStop}%
\bibitem [{\citenamefont {Zhang}\ \emph {et~al.}(2000)\citenamefont {Zhang},
  \citenamefont {Ong}, \citenamefont {Xu}, \citenamefont {Krishana},
  \citenamefont {Gagnon},\ and\ \citenamefont {Taillefer}}]{Zhang2000}%
  \BibitemOpen
  \bibfield  {author} {\bibinfo {author} {\bibfnamefont {Y.}~\bibnamefont
  {Zhang}}, \bibinfo {author} {\bibfnamefont {N.~P.}\ \bibnamefont {Ong}},
  \bibinfo {author} {\bibfnamefont {Z.~A.}\ \bibnamefont {Xu}}, \bibinfo
  {author} {\bibfnamefont {K.}~\bibnamefont {Krishana}}, \bibinfo {author}
  {\bibfnamefont {R.}~\bibnamefont {Gagnon}},\ and\ \bibinfo {author}
  {\bibfnamefont {L.}~\bibnamefont {Taillefer}},\ }\bibfield  {title} {\bibinfo
  {title} {Determining the wiedemann-franz ratio from the thermal hall
  conductivity: Application to cu and
  ${\mathrm{yba}}_{2}{\mathrm{cu}}_{3}{O}_{6.95}$},\ }\href
  {https://doi.org/10.1103/PhysRevLett.84.2219} {\bibfield  {journal} {\bibinfo
   {journal} {Phys. Rev. Lett.}\ }\textbf {\bibinfo {volume} {84}},\ \bibinfo
  {pages} {2219} (\bibinfo {year} {2000})}\BibitemShut {NoStop}%
\bibitem [{\citenamefont {Behnia}(2015{\natexlab{a}})}]{Behnia2015b}%
  \BibitemOpen
  \bibfield  {author} {\bibinfo {author} {\bibfnamefont {K.}~\bibnamefont
  {Behnia}},\ }\href@noop {} {\emph {\bibinfo {title} {{Fundamentals of
  Thermoelectricity}}}}\ (\bibinfo  {publisher} {Oxford University Press},\
  \bibinfo {year} {2015})\BibitemShut {NoStop}%
\bibitem [{SM()}]{SM}%
  \BibitemOpen
  \href@noop {} {\bibinfo {title} {See the supplementary material for more
  details on the samples and measurement methods.}}\BibitemShut {Stop}%
\bibitem [{\citenamefont {Collignon}\ \emph {et~al.}(2021)\citenamefont
  {Collignon}, \citenamefont {Awashima}, \citenamefont {Ravi}, \citenamefont
  {Lin}, \citenamefont {Rischau}, \citenamefont {Acheche}, \citenamefont
  {Vignolle}, \citenamefont {Proust}, \citenamefont {Fuseya}, \citenamefont
  {Behnia},\ and\ \citenamefont {Fauqu\'e}}]{Collignon2021}%
  \BibitemOpen
  \bibfield  {author} {\bibinfo {author} {\bibfnamefont {C.}~\bibnamefont
  {Collignon}}, \bibinfo {author} {\bibfnamefont {Y.}~\bibnamefont {Awashima}},
  \bibinfo {author} {\bibnamefont {Ravi}}, \bibinfo {author} {\bibfnamefont
  {X.}~\bibnamefont {Lin}}, \bibinfo {author} {\bibfnamefont {C.~W.}\
  \bibnamefont {Rischau}}, \bibinfo {author} {\bibfnamefont {A.}~\bibnamefont
  {Acheche}}, \bibinfo {author} {\bibfnamefont {B.}~\bibnamefont {Vignolle}},
  \bibinfo {author} {\bibfnamefont {C.}~\bibnamefont {Proust}}, \bibinfo
  {author} {\bibfnamefont {Y.}~\bibnamefont {Fuseya}}, \bibinfo {author}
  {\bibfnamefont {K.}~\bibnamefont {Behnia}},\ and\ \bibinfo {author}
  {\bibfnamefont {B.}~\bibnamefont {Fauqu\'e}},\ }\bibfield  {title} {\bibinfo
  {title} {Quasi-isotropic orbital magnetoresistance in lightly doped
  $\mathrm{SrTiO_{3}}$},\ }\href
  {https://doi.org/10.1103/PhysRevMaterials.5.065002} {\bibfield  {journal}
  {\bibinfo  {journal} {Phys. Rev. Materials}\ }\textbf {\bibinfo {volume}
  {5}},\ \bibinfo {pages} {065002} (\bibinfo {year} {2021})}\BibitemShut
  {NoStop}%
\bibitem [{\citenamefont {Cain}\ \emph {et~al.}(2013)\citenamefont {Cain},
  \citenamefont {Kajdos},\ and\ \citenamefont {Stemmer}}]{Cain2013}%
  \BibitemOpen
  \bibfield  {author} {\bibinfo {author} {\bibfnamefont {T.~A.}\ \bibnamefont
  {Cain}}, \bibinfo {author} {\bibfnamefont {A.~P.}\ \bibnamefont {Kajdos}},\
  and\ \bibinfo {author} {\bibfnamefont {S.}~\bibnamefont {Stemmer}},\
  }\bibfield  {title} {\bibinfo {title} {La-doped {SrTiO$_3$} films with large
  cryogenic thermoelectric power factors},\ }\href@noop {} {\bibfield
  {journal} {\bibinfo  {journal} {Applied Physics Letters}\ }\textbf {\bibinfo
  {volume} {102}},\ \bibinfo {pages} {182101} (\bibinfo {year}
  {2013})}\BibitemShut {NoStop}%
\bibitem [{\citenamefont {Collignon}\ \emph {et~al.}(2020)\citenamefont
  {Collignon}, \citenamefont {Bourges}, \citenamefont {Fauqu\'e},\ and\
  \citenamefont {Behnia}}]{Collignon2020}%
  \BibitemOpen
  \bibfield  {author} {\bibinfo {author} {\bibfnamefont {C.}~\bibnamefont
  {Collignon}}, \bibinfo {author} {\bibfnamefont {P.}~\bibnamefont {Bourges}},
  \bibinfo {author} {\bibfnamefont {B.}~\bibnamefont {Fauqu\'e}},\ and\
  \bibinfo {author} {\bibfnamefont {K.}~\bibnamefont {Behnia}},\ }\bibfield
  {title} {\bibinfo {title} {Heavy nondegenerate electrons in doped strontium
  titanate},\ }\href {https://doi.org/10.1103/PhysRevX.10.031025} {\bibfield
  {journal} {\bibinfo  {journal} {Phys. Rev. X}\ }\textbf {\bibinfo {volume}
  {10}},\ \bibinfo {pages} {031025} (\bibinfo {year} {2020})}\BibitemShut
  {NoStop}%
\bibitem [{\citenamefont {Mousatov}\ and\ \citenamefont
  {Hartnoll}(2021)}]{Mousatov2021}%
  \BibitemOpen
  \bibfield  {author} {\bibinfo {author} {\bibfnamefont {C.~H.}\ \bibnamefont
  {Mousatov}}\ and\ \bibinfo {author} {\bibfnamefont {S.~A.}\ \bibnamefont
  {Hartnoll}},\ }\bibfield  {title} {\bibinfo {title} {Phonons, electrons and
  thermal transport in planckian high tc materials},\ }\href@noop {} {\bibfield
   {journal} {\bibinfo  {journal} {npj Quantum Materials}\ }\textbf {\bibinfo
  {volume} {6}},\ \bibinfo {pages} {81} (\bibinfo {year} {2021})}\BibitemShut
  {NoStop}%
\bibitem [{\citenamefont {Rehwald}(1970)}]{REHWALD}%
  \BibitemOpen
  \bibfield  {author} {\bibinfo {author} {\bibfnamefont {W.}~\bibnamefont
  {Rehwald}},\ }\bibfield  {title} {\bibinfo {title} {Anomalous ultrasonic
  attenuation at the 105°k transition in strontium titanate},\ }\href
  {https://doi.org/https://doi.org/10.1016/0038-1098(70)90159-6} {\bibfield
  {journal} {\bibinfo  {journal} {Solid State Communications}\ }\textbf
  {\bibinfo {volume} {8}},\ \bibinfo {pages} {607} (\bibinfo {year}
  {1970})}\BibitemShut {NoStop}%
\bibitem [{\citenamefont {Kumar}\ \emph {et~al.}(2021)\citenamefont {Kumar},
  \citenamefont {Yudson},\ and\ \citenamefont {Maslov}}]{Kumar2021}%
  \BibitemOpen
  \bibfield  {author} {\bibinfo {author} {\bibfnamefont {A.}~\bibnamefont
  {Kumar}}, \bibinfo {author} {\bibfnamefont {V.~I.}\ \bibnamefont {Yudson}},\
  and\ \bibinfo {author} {\bibfnamefont {D.~L.}\ \bibnamefont {Maslov}},\
  }\bibfield  {title} {\bibinfo {title} {Quasiparticle and nonquasiparticle
  transport in doped quantum paraelectrics},\ }\href
  {https://doi.org/10.1103/PhysRevLett.126.076601} {\bibfield  {journal}
  {\bibinfo  {journal} {Phys. Rev. Lett.}\ }\textbf {\bibinfo {volume} {126}},\
  \bibinfo {pages} {076601} (\bibinfo {year} {2021})}\BibitemShut {NoStop}%
\bibitem [{\citenamefont {Nazaryan}\ and\ \citenamefont
  {Feigel'man}(2021)}]{Nazaryan2021}%
  \BibitemOpen
  \bibfield  {author} {\bibinfo {author} {\bibfnamefont {K.~G.}\ \bibnamefont
  {Nazaryan}}\ and\ \bibinfo {author} {\bibfnamefont {M.~V.}\ \bibnamefont
  {Feigel'man}},\ }\bibfield  {title} {\bibinfo {title} {Conductivity and
  thermoelectric coefficients of doped {${\mathrm{SrTiO}}_{3}$} at high
  temperatures},\ }\href {https://doi.org/10.1103/PhysRevB.104.115201}
  {\bibfield  {journal} {\bibinfo  {journal} {Phys. Rev. B}\ }\textbf {\bibinfo
  {volume} {104}},\ \bibinfo {pages} {115201} (\bibinfo {year}
  {2021})}\BibitemShut {NoStop}%
\bibitem [{\citenamefont {Lin}\ \emph {et~al.}(2015)\citenamefont {Lin},
  \citenamefont {Fauqu{\'e}},\ and\ \citenamefont {Behnia}}]{Lin2015}%
  \BibitemOpen
  \bibfield  {author} {\bibinfo {author} {\bibfnamefont {X.}~\bibnamefont
  {Lin}}, \bibinfo {author} {\bibfnamefont {B.}~\bibnamefont {Fauqu{\'e}}},\
  and\ \bibinfo {author} {\bibfnamefont {K.}~\bibnamefont {Behnia}},\
  }\bibfield  {title} {\bibinfo {title} {{Scalable T$^2$ resistivity in a small
  single-component Fermi surface}},\ }\href
  {https://doi.org/10.1126/science.aaa8655} {\bibfield  {journal} {\bibinfo
  {journal} {Science}\ }\textbf {\bibinfo {volume} {349}},\ \bibinfo {pages}
  {945} (\bibinfo {year} {2015})}\BibitemShut {NoStop}%
\bibitem [{\citenamefont {Behnia}\ and\ \citenamefont
  {Aubin}(2016)}]{Behnia_2016}%
  \BibitemOpen
  \bibfield  {author} {\bibinfo {author} {\bibfnamefont {K.}~\bibnamefont
  {Behnia}}\ and\ \bibinfo {author} {\bibfnamefont {H.}~\bibnamefont {Aubin}},\
  }\bibfield  {title} {\bibinfo {title} {Nernst effect in metals and
  superconductors: a review of concepts and experiments},\ }\href
  {https://doi.org/10.1088/0034-4885/79/4/046502} {\bibfield  {journal}
  {\bibinfo  {journal} {Reports on Progress in Physics}\ }\textbf {\bibinfo
  {volume} {79}},\ \bibinfo {pages} {046502} (\bibinfo {year}
  {2016})}\BibitemShut {NoStop}%
\bibitem [{\citenamefont {Behnia}(2015{\natexlab{b}})}]{Behnia_2015}%
  \BibitemOpen
  \bibfield  {author} {\bibinfo {author} {\bibfnamefont {K.}~\bibnamefont
  {Behnia}},\ }\bibfield  {title} {\bibinfo {title} {On mobility of electrons
  in a shallow fermi sea over a rough seafloor},\ }\href
  {https://doi.org/10.1088/0953-8984/27/37/375501} {\bibfield  {journal}
  {\bibinfo  {journal} {Journal of Physics: Condensed Matter}\ }\textbf
  {\bibinfo {volume} {27}},\ \bibinfo {pages} {375501} (\bibinfo {year}
  {2015}{\natexlab{b}})}\BibitemShut {NoStop}%
\bibitem [{\citenamefont {Fauqu\'e}\ \emph {et~al.}(2013)\citenamefont
  {Fauqu\'e}, \citenamefont {Butch}, \citenamefont {Syers}, \citenamefont
  {Paglione}, \citenamefont {Wiedmann}, \citenamefont {Collaudin},
  \citenamefont {Grena}, \citenamefont {Zeitler},\ and\ \citenamefont
  {Behnia}}]{Fauque2013}%
  \BibitemOpen
  \bibfield  {author} {\bibinfo {author} {\bibfnamefont {B.}~\bibnamefont
  {Fauqu\'e}}, \bibinfo {author} {\bibfnamefont {N.~P.}\ \bibnamefont {Butch}},
  \bibinfo {author} {\bibfnamefont {P.}~\bibnamefont {Syers}}, \bibinfo
  {author} {\bibfnamefont {J.}~\bibnamefont {Paglione}}, \bibinfo {author}
  {\bibfnamefont {S.}~\bibnamefont {Wiedmann}}, \bibinfo {author}
  {\bibfnamefont {A.}~\bibnamefont {Collaudin}}, \bibinfo {author}
  {\bibfnamefont {B.}~\bibnamefont {Grena}}, \bibinfo {author} {\bibfnamefont
  {U.}~\bibnamefont {Zeitler}},\ and\ \bibinfo {author} {\bibfnamefont
  {K.}~\bibnamefont {Behnia}},\ }\bibfield  {title} {\bibinfo {title}
  {Magnetothermoelectric properties of $\mathrm{Bi_{2}Se_{3}}$},\ }\href
  {https://doi.org/10.1103/PhysRevB.87.035133} {\bibfield  {journal} {\bibinfo
  {journal} {Phys. Rev. B}\ }\textbf {\bibinfo {volume} {87}},\ \bibinfo
  {pages} {035133} (\bibinfo {year} {2013})}\BibitemShut {NoStop}%
\bibitem [{\citenamefont {Jaoui}\ \emph {et~al.}(2020)\citenamefont {Jaoui},
  \citenamefont {Seyfarth}, \citenamefont {Rischau}, \citenamefont {Wiedmann},
  \citenamefont {Benhabib}, \citenamefont {Proust}, \citenamefont {Behnia},\
  and\ \citenamefont {Fauqu{\'e}}}]{Jaoui2020}%
  \BibitemOpen
  \bibfield  {author} {\bibinfo {author} {\bibfnamefont {A.}~\bibnamefont
  {Jaoui}}, \bibinfo {author} {\bibfnamefont {G.}~\bibnamefont {Seyfarth}},
  \bibinfo {author} {\bibfnamefont {C.~W.}\ \bibnamefont {Rischau}}, \bibinfo
  {author} {\bibfnamefont {S.}~\bibnamefont {Wiedmann}}, \bibinfo {author}
  {\bibfnamefont {S.}~\bibnamefont {Benhabib}}, \bibinfo {author}
  {\bibfnamefont {C.}~\bibnamefont {Proust}}, \bibinfo {author} {\bibfnamefont
  {K.}~\bibnamefont {Behnia}},\ and\ \bibinfo {author} {\bibfnamefont
  {B.}~\bibnamefont {Fauqu{\'e}}},\ }\bibfield  {title} {\bibinfo {title}
  {Giant seebeck effect across the field-induced metal-insulator transition of
  inas},\ }\href {https://doi.org/10.1038/s41535-020-00296-0} {\bibfield
  {journal} {\bibinfo  {journal} {npj Quantum Materials}\ }\textbf {\bibinfo
  {volume} {5}},\ \bibinfo {pages} {94} (\bibinfo {year} {2020})}\BibitemShut
  {NoStop}%
\bibitem [{\citenamefont {Ravich}\ \emph {et~al.}(1970)\citenamefont {Ravich},
  \citenamefont {Efimova},\ and\ \citenamefont {Smirnov}}]{bookPbTe}%
  \BibitemOpen
  \bibfield  {author} {\bibinfo {author} {\bibfnamefont {Y.~I.}\ \bibnamefont
  {Ravich}}, \bibinfo {author} {\bibfnamefont {B.~A.}\ \bibnamefont
  {Efimova}},\ and\ \bibinfo {author} {\bibfnamefont {I.~A.}\ \bibnamefont
  {Smirnov}},\ }\href@noop {} {\emph {\bibinfo {title} {Semiconducting Lead
  Chalcogenides}}},\ \bibinfo {edition} {1st}\ ed.,\ Monographs in
  Semiconductor Physics 5\ (\bibinfo  {publisher} {Springer US},\ \bibinfo
  {year} {1970})\BibitemShut {NoStop}%
\bibitem [{\citenamefont {Ando}\ \emph {et~al.}(2004)\citenamefont {Ando},
  \citenamefont {Kurita}, \citenamefont {Komiya}, \citenamefont {Ono},\ and\
  \citenamefont {Segawa}}]{Ando2004}%
  \BibitemOpen
  \bibfield  {author} {\bibinfo {author} {\bibfnamefont {Y.}~\bibnamefont
  {Ando}}, \bibinfo {author} {\bibfnamefont {Y.}~\bibnamefont {Kurita}},
  \bibinfo {author} {\bibfnamefont {S.}~\bibnamefont {Komiya}}, \bibinfo
  {author} {\bibfnamefont {S.}~\bibnamefont {Ono}},\ and\ \bibinfo {author}
  {\bibfnamefont {K.}~\bibnamefont {Segawa}},\ }\bibfield  {title} {\bibinfo
  {title} {Evolution of the {Hall} coefficient and the peculiar electronic
  structure of the cuprate superconductors},\ }\href
  {https://doi.org/10.1103/PhysRevLett.92.197001} {\bibfield  {journal}
  {\bibinfo  {journal} {Phys. Rev. Lett.}\ }\textbf {\bibinfo {volume} {92}},\
  \bibinfo {pages} {197001} (\bibinfo {year} {2004})}\BibitemShut {NoStop}%
\bibitem [{\citenamefont {Spinelli}\ \emph {et~al.}(2010)\citenamefont
  {Spinelli}, \citenamefont {Torija}, \citenamefont {Liu}, \citenamefont
  {Jan},\ and\ \citenamefont {Leighton}}]{Spinelli2010}%
  \BibitemOpen
  \bibfield  {author} {\bibinfo {author} {\bibfnamefont {A.}~\bibnamefont
  {Spinelli}}, \bibinfo {author} {\bibfnamefont {M.~A.}\ \bibnamefont
  {Torija}}, \bibinfo {author} {\bibfnamefont {C.}~\bibnamefont {Liu}},
  \bibinfo {author} {\bibfnamefont {C.}~\bibnamefont {Jan}},\ and\ \bibinfo
  {author} {\bibfnamefont {C.}~\bibnamefont {Leighton}},\ }\bibfield  {title}
  {\bibinfo {title} {Electronic transport in doped ${\text{srtio}}_{3}$:
  Conduction mechanisms and potential applications},\ }\href
  {https://doi.org/10.1103/PhysRevB.81.155110} {\bibfield  {journal} {\bibinfo
  {journal} {Phys. Rev. B}\ }\textbf {\bibinfo {volume} {81}},\ \bibinfo
  {pages} {155110} (\bibinfo {year} {2010})}\BibitemShut {NoStop}%
\bibitem [{\citenamefont {de~Lima}\ \emph {et~al.}(2015)\citenamefont
  {de~Lima}, \citenamefont {da~Luz}, \citenamefont {Oliveira}, \citenamefont
  {Alves}, \citenamefont {dos Santos}, \citenamefont {Jomard}, \citenamefont
  {Sidis}, \citenamefont {Bourges}, \citenamefont {Harms}, \citenamefont
  {Grams}, \citenamefont {Hemberger}, \citenamefont {Lin}, \citenamefont
  {Fauqu\'e},\ and\ \citenamefont {Behnia}}]{delima2015}%
  \BibitemOpen
  \bibfield  {author} {\bibinfo {author} {\bibfnamefont {B.~S.}\ \bibnamefont
  {de~Lima}}, \bibinfo {author} {\bibfnamefont {M.~S.}\ \bibnamefont {da~Luz}},
  \bibinfo {author} {\bibfnamefont {F.~S.}\ \bibnamefont {Oliveira}}, \bibinfo
  {author} {\bibfnamefont {L.~M.~S.}\ \bibnamefont {Alves}}, \bibinfo {author}
  {\bibfnamefont {C.~A.~M.}\ \bibnamefont {dos Santos}}, \bibinfo {author}
  {\bibfnamefont {F.}~\bibnamefont {Jomard}}, \bibinfo {author} {\bibfnamefont
  {Y.}~\bibnamefont {Sidis}}, \bibinfo {author} {\bibfnamefont
  {P.}~\bibnamefont {Bourges}}, \bibinfo {author} {\bibfnamefont
  {S.}~\bibnamefont {Harms}}, \bibinfo {author} {\bibfnamefont {C.~P.}\
  \bibnamefont {Grams}}, \bibinfo {author} {\bibfnamefont {J.}~\bibnamefont
  {Hemberger}}, \bibinfo {author} {\bibfnamefont {X.}~\bibnamefont {Lin}},
  \bibinfo {author} {\bibfnamefont {B.}~\bibnamefont {Fauqu\'e}},\ and\
  \bibinfo {author} {\bibfnamefont {K.}~\bibnamefont {Behnia}},\ }\bibfield
  {title} {\bibinfo {title} {Interplay between antiferrodistortive,
  ferroelectric, and superconducting instabilities in
  $\mathrm{Sr_{1-x}Ca_{x}TiO_{3}}$},\ }\href
  {https://doi.org/10.1103/PhysRevB.91.045108} {\bibfield  {journal} {\bibinfo
  {journal} {Phys. Rev. B}\ }\textbf {\bibinfo {volume} {91}},\ \bibinfo
  {pages} {045108} (\bibinfo {year} {2015})}\BibitemShut {NoStop}%
\bibitem [{\citenamefont {Coey}\ \emph {et~al.}(2016)\citenamefont {Coey},
  \citenamefont {Venkatesan},\ and\ \citenamefont {Stamenov}}]{Coey_2016}%
  \BibitemOpen
  \bibfield  {author} {\bibinfo {author} {\bibfnamefont {J.~M.~D.}\
  \bibnamefont {Coey}}, \bibinfo {author} {\bibfnamefont {M.}~\bibnamefont
  {Venkatesan}},\ and\ \bibinfo {author} {\bibfnamefont {P.}~\bibnamefont
  {Stamenov}},\ }\bibfield  {title} {\bibinfo {title} {Surface magnetism of
  strontium titanate},\ }\href {https://doi.org/10.1088/0953-8984/28/48/485001}
  {\bibfield  {journal} {\bibinfo  {journal} {Journal of Physics: Condensed
  Matter}\ }\textbf {\bibinfo {volume} {28}},\ \bibinfo {pages} {485001}
  (\bibinfo {year} {2016})}\BibitemShut {NoStop}%
\end{thebibliography}%
\appendix
\onecolumngrid
\renewcommand{\thesection}{S\arabic{section}}
\renewcommand{\thetable}{S\arabic{table}}
\renewcommand{\thefigure}{S\arabic{figure}}
\renewcommand{\theequation}{S\arabic{equation}}
\setcounter{section}{0}
\setcounter{figure}{0}
\setcounter{table}{0}
\setcounter{equation}{0}
\begin{center}{\large\bf Supplemental Material for ``Phonon drag  thermal Hall effect in metallic strontium titanate"}\\
\end{center}
\setcounter{figure}{0}

\section{Samples}
\begin{figure}[h]
\centering
\includegraphics[width=10cm]{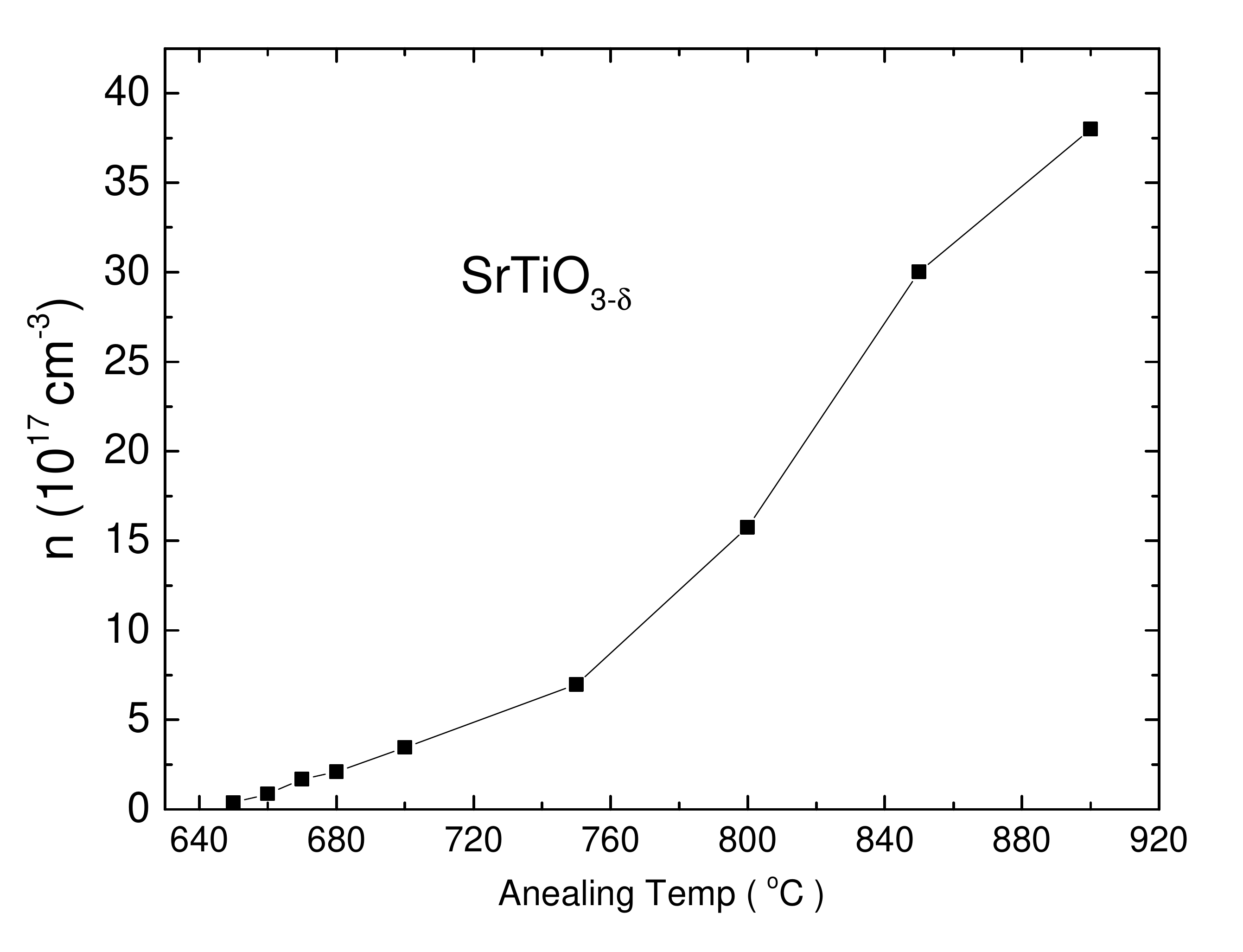}
    \caption{ Carrier density of SrTiO$_{3-\delta}$ as a function of annealing temperature (the vacuum chamber was kept below $10^{-6}$ mbar).}
\label{fig: Sample Preparation}\
\end{figure}

SrTiO$_3$ and Ca-doped SrTiO$_3$ single crystals were provided by CrysTech and SurfaceNet GmbH. Oxygen-deficient SrTiO$_3$ samples were obtained by annealing them in a temperature range extending from 700$^oC$ to  1000 $^oC$ under a high vacuum ($< 10^{-6}$ mbar) for 1 to 2 hours. Fig \ref{fig: Sample Preparation} shows the Hall carrier density of the oxygen-reduced samples as function of the annealing temperature. These results are similar to what was previously reported in oxygen reduced SrTiO$_3$ samples \cite{Spinelli2010,Lin2013}.

\begin{table}[htb]
\begin{tabular}{c|c|c|c|c|c}
\hline
 No. & Supplier & Dimensions (mm) & Peak of $\kappa_{xx}$ (W.K$^{-1}$m$^{-1}$) & Peak of $\kappa_{xy}$ (mW.K$^{-1}$.m$^{-1}$) & Ca concentration (ppm)  \\
\hline
Ref\#1 \cite{Li2020} & SurfaceNet  & $5\times5\times0.5$ & 37 & 80 & 3.8 \\
\hline
Ref\#3 \cite{Li2020} & SurfaceNet  & $5\times5\times0.5$ & 35 & 26 & 2.5 \\
\hline
\#1  & CrysTech  & $5\times5\times0.93$ & 38.6 & 21.6 & 290 \\
\hline
\#2  & SurfaceNet  & $5\times5\times0.5$ & 34 & 24 & 69 \\
\hline
\#3  & SurfaceNet  & $5\times5\times0.5$ & 32 & 24 & - \\
\hline
\#4  & CrysTech  & $5\times5\times0.82$ & - & 3.3 & 4500 \\
\hline
\#5  & CrysTech  & $5\times5\times0.99$ & - & 0.6 & 9100 \\
\hline

\end{tabular}
\caption{Properties of the insulating samples studied in this work. Ca concentration was determined using Secondary Ion Mass Spectrometry (SIMS) and is expressed in part per million (ppm).}
\label{table-Properties of STO}
\end{table}
 
 The list of the pristine SrTiO$_3$ samples studied in this work, together with their dimensions, the peak amplitude of $\kappa_{xx}$ and $\kappa_{xy}$ and their Ca concentration, is shown in Table \ref{table-Properties of STO}. Ca concentration has been determined using Secondary Ion Mass Spectrometry (SIMS) like in \cite{delima2015}.
 
 The amplitude of the $\kappa_{xx}$ peak is similar in all pristine samples. On the other hand, we found a significant difference in the amplitude of $\kappa_{xy}$ peak from batch to batch. This variation does not clearly correlate with the amount of residual Ca impurities  determined by SIMS. Nominally pristine strontium titanate crystals are known to host other types of impurities such as Fe, O, Ni with a typical concentration of 5 (ppm) \cite{Coey_2016}. Further studies may find a correlation between the amplitude of the $\kappa_{xy}$ peak and the concentration of such impurities in nominally pure SrTiO$_3$ samples. In this study all the reduced samples have been prepared from the same batch \#3 . 
 
\section{Electric and thermoelectric transport in $SrTiO_{3-\delta}$ samples }

\begin{figure}[h]
\centering
\includegraphics[width=14cm]{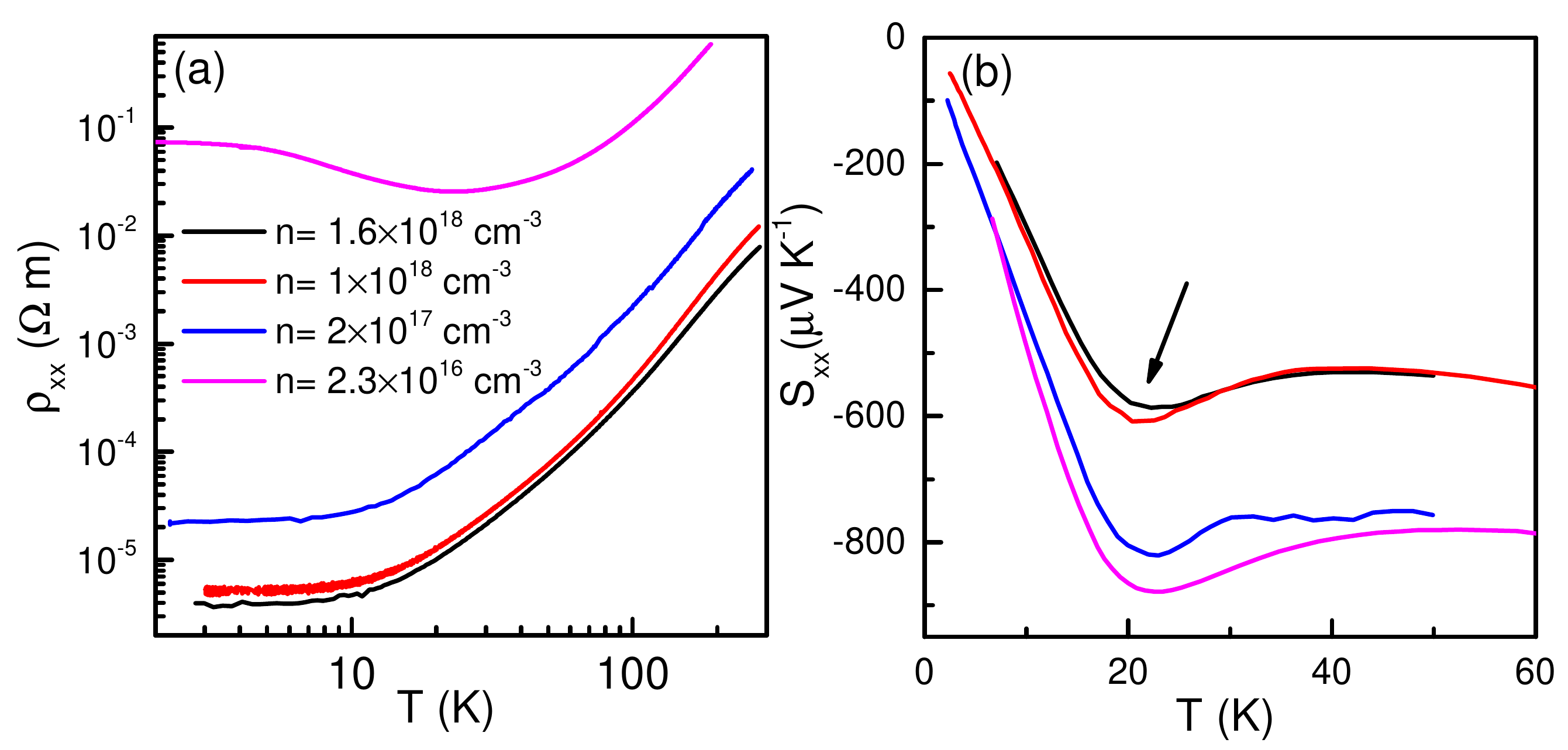}
    \caption{ (a) Resistivity ($\rho_{xx}$) and (b) Seebeck coefficient ($S_{xx}$) as function of temperature for SrTiO$_{3-\delta}$ samples with different carrier concentrations. The arrow in (b) indicates the phonon drag peak. }
\label{fig: RHo and Seebeck}
\end{figure}

Fig.\ref{fig: RHo and Seebeck} (a) shows the temperature dependence of the resistivity ($\rho_{xx}$) in SrTiO$_{3-\delta}$ with Hall carrier density ($n$) ranging from 2.3$\times$10$^{16}$ to 1.6$\times$10$^{18}$cm$^{-3}$, in good agreement with previous works \cite{Spinelli2010,Lin2015}. Samples with carrier density larger than 10$^{17}$cm$^{-3}$ display a metallic behaviour down to the lowest temperature. In contrast, the lowest doped sample ($n=2.3\times10^{16}cm^{-3}$) shows an upturn below  30 K, possibly due to inhomogeneous oxygen vacancies distribution or the approach of a genuine metal-insulator transition. This is in agreement with a previous study, which reported an insulating behavior in samples with a carrier density below $n=3\times10^{16}cm^{-3}$ \cite{Spinelli2010}.

\begin{figure}[h]
\centering
\includegraphics[width=17cm]{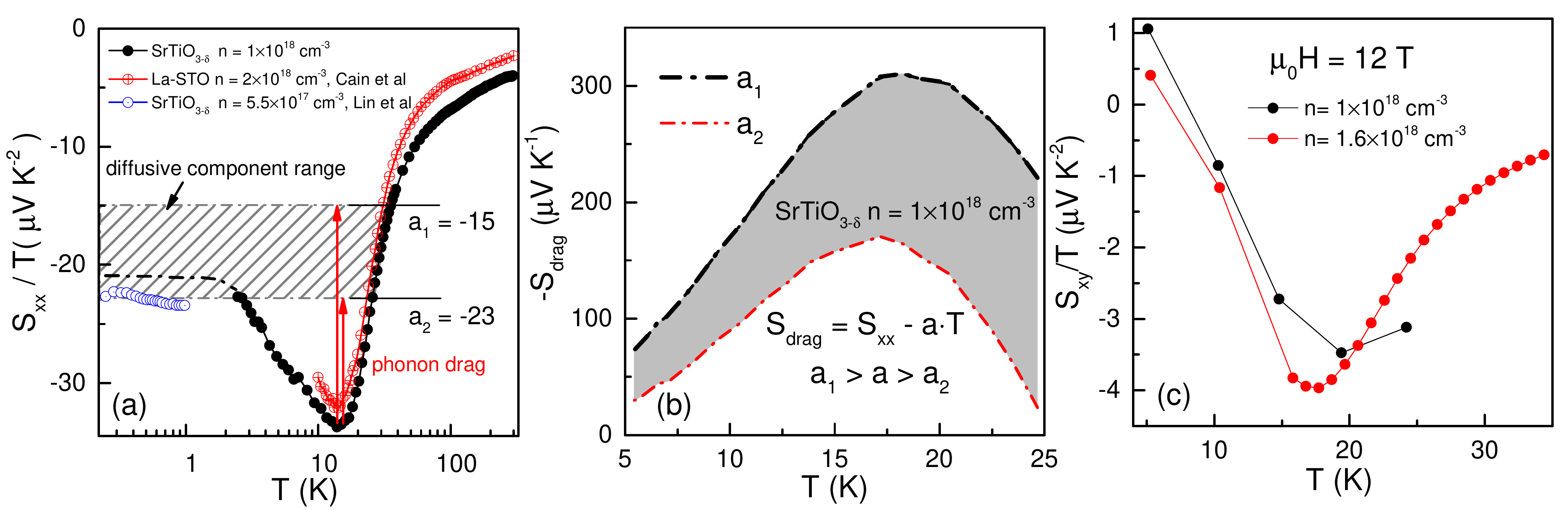}
    \caption{(a) and (b) $\rho_{xx}$ and $\rho_{xy}$ as function of the magnetic field (B) for T= 5.3 K to 29.8 K.(c) Hall angle ($\frac{\rho_{xy}}{\rho_{xx}}$) $vs.$ B.(d) Hall coefficient $R_H = \frac{\rho_{xy}}{B}=\frac{1}{ne}$ $vs.$ B.(e) Electron mobility ($\mu = \frac{1}{B}\cdot\frac{\rho_{xy}}{\rho_{xx}}$) $vs.$ B.(f) the mean free path of electrons, $l=\tau\cdot v = \frac{\sigma\cdot m^*}{ne^2}\cdot c$ ($m^* = 1.8 m_e$ \cite{Lin2013} and $c=3\times 10^8 m/s $)  $vs.$ B }
\label{fig: electric properties}
\end{figure}

Fig. \ref{fig: RHo and Seebeck} (b) shows the temperature dependence of the Seebeck coefficient ($S_{xx}=\frac{E_x}{\Delta_x T}$) in our SrTiO$_{3-\delta}$ samples. $S_{xx}$ peaks around 20 $K$. This is a signature of phonon drag as discussed below.  

Fig. \ref{fig: electric properties} shows the electric properties of $n = 1.6\times 10^{18}$ sample. The carrier density is determined  by measuring the Hall coefficient. The data is similar to what was reported in ref. \cite{Collignon2021}.

\section{Diffusive and phonon drag components of the thermoelectric response }

\begin{figure}[h]
\centering
\includegraphics[width=17cm]{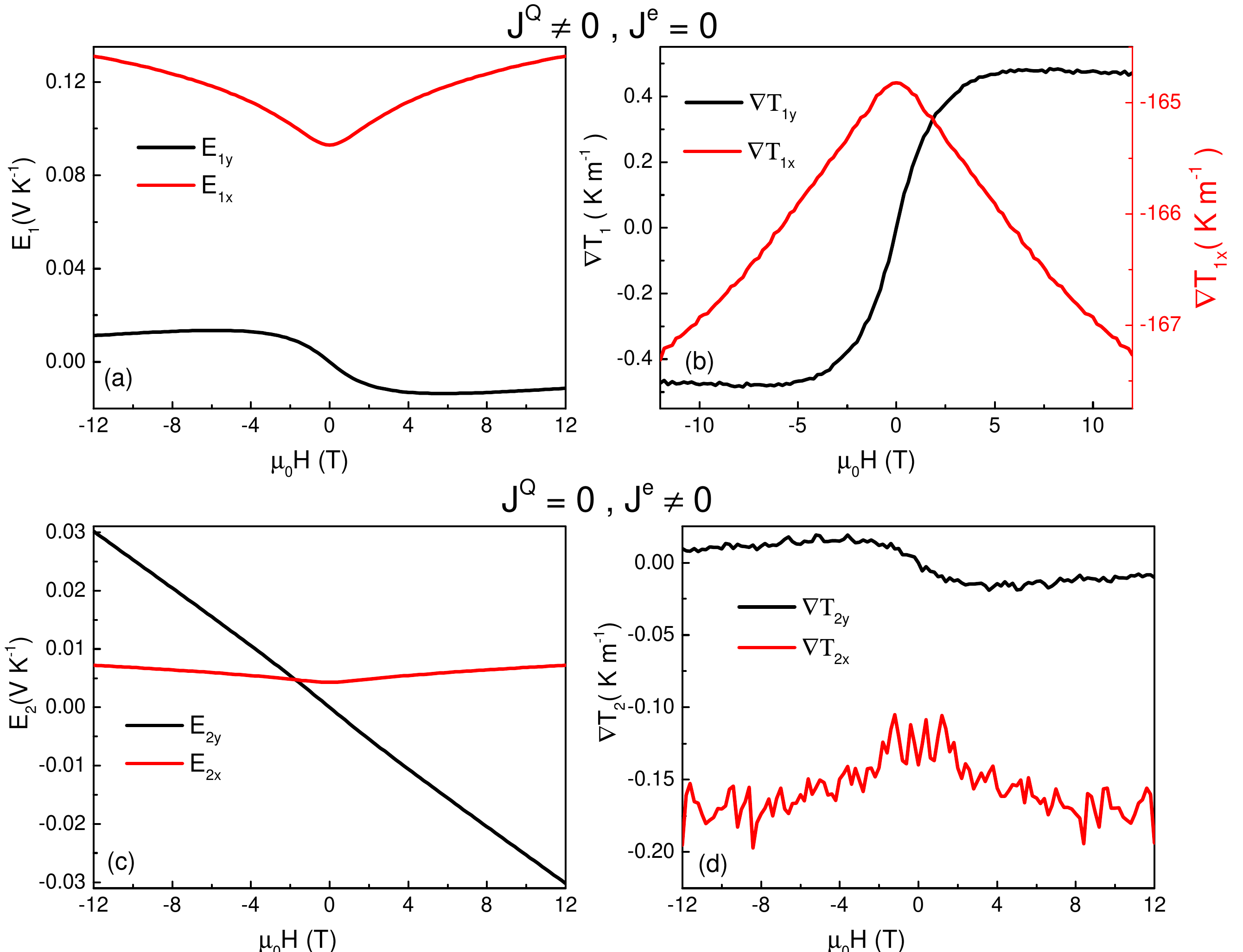}
    \caption{ (a) Temperature dependence of $S_{xx}/T$ in SrTiO$_{3-\delta}$  for n=1$\times$ 10$^{18}$ cm$^{-3}$ compared with previous works \cite{Cain2013,Lin2013}. The hatched area is the estimated diffusive contribution  using Eq.\ref{S1}. (b) Amplitude of the phonon drag contribution of S$_{xx}$ (gray area) as function of the temperature. $S_{drag}=S_{xx}-a\cdot T$ where  $a\cdot T$ is the diffusive contribution shown in (a).(c) Temperature dependence of  $S_{xy}/T$ at $B$= 12 T for n=1$\times$ 10$^{18}$ cm$^{-3}$ and 1.6$\times$ 10$^{18}$ cm$^{-3}$. A phonon drag contribution leads to a peaks in $S_{xy}$ around 20 K. }
\label{fig: thermopower}
\end{figure}

Fig. \ref{fig: thermopower} (a) compares the temperature dependence of $S_{xx}/T$ in doped sample ($n$ = 1 $\times$ 10$^{18}$ cm$^{-3}$) with previous measurements \cite{Cain2013,Lin2013}. Sub-kelvin measurements have shown that $S_{xx}/T$ becomes flat at low temperature (see Fig. \ref{fig: thermopower} (a)) with the expected amplitude for the diffusive response of a degenerate semiconductor with an energy-independent mean free path \cite{Lin2013} :

\begin{equation}\label{S1}
 \lvert \frac{S}{T} \rvert = \frac{\pi^2}{3} \frac{k_B}{e} \frac{1}{T_F}
\end{equation}

$T_F$ is the Fermi temperature, which can be deduced from quantum oscillations. Using Eq.\ref{S1} we estimate the amplitude of the diffusive response  in our sample with $n=1\times10^{18}cm^{-3}$ to be in the range of $-15\sim-23\ \mu V K^{-2} $. On top of it, there is a second contribution, the phonon drag (labelled $S_{drag}$), shown in gray  in Fig.\ref{fig: thermopower}b). $S_{drag}$ peaks  around 20 $K$, in good agreement with previous work \cite{Cain2013}. A similar phonon drag contribution is also detected in the Nernst effect ($S_{xy}=\frac{E_y}{\Delta T_x}$) see Fig. \ref{fig: thermopower} (c). 

\section{Verifying Onsager reciprocity}

According to the Ohm's law, the electrical field $\Vec{E}$ and the charge flux density ($\Vec{J^e}$) are linked through the electric conductivity $\overline{\sigma}$ : 

\begin{equation}\label{S2}
 \Vec{J^e}=\overline{\sigma} \Vec{E}
\end{equation}

Similarly, according to the Fourier's law, the thermal conductivity $\overline{\kappa} $ links the thermal gradient $\Vec{\nabla T}$  and the heat flux density $\Vec{J^Q}$:

\begin{equation}\label{S3}
 \Vec{J^Q}=-\overline{\kappa} \Vec{\nabla T}  
\end{equation}

The existence of the thermoelectric response modifies both equations: 
\begin{equation}\label{S4}
 \Vec{J^e}=\overline{\sigma'} \Vec{E}-\overline{\alpha} \Vec{\nabla T}
\end{equation}

\begin{equation}\label{S5}
\ \Vec{J^Q}=\overline{\alpha} T \Vec{E}-\overline{\kappa'} \Vec{\nabla T}  
\end{equation}

Here $\overline{\sigma'}$, $\overline{\kappa'}$ and $\overline{\alpha} $  are the electric, thermal  and thermoelectric conductivity tensors. The thermoelectric contribution in Eq.\ref{S4} ($\alpha \nabla T$) and in Eq.\ref{S5} ($\alpha  T E $ ) render $\overline{\sigma}$ and $\overline{\kappa}$ different from $\overline{\sigma'}$ and $\overline{\kappa'}$. In most cases, this contribution to $\overline{\kappa'}$ is negligibly small. However, in our case, this is not true for the transverse component of $\overline{\kappa}$.  In other words, $\kappa_{xy}\neq \kappa^{\prime}_{xy}$. 

\begin{figure}[h]
\centering
\includegraphics[width=17cm]{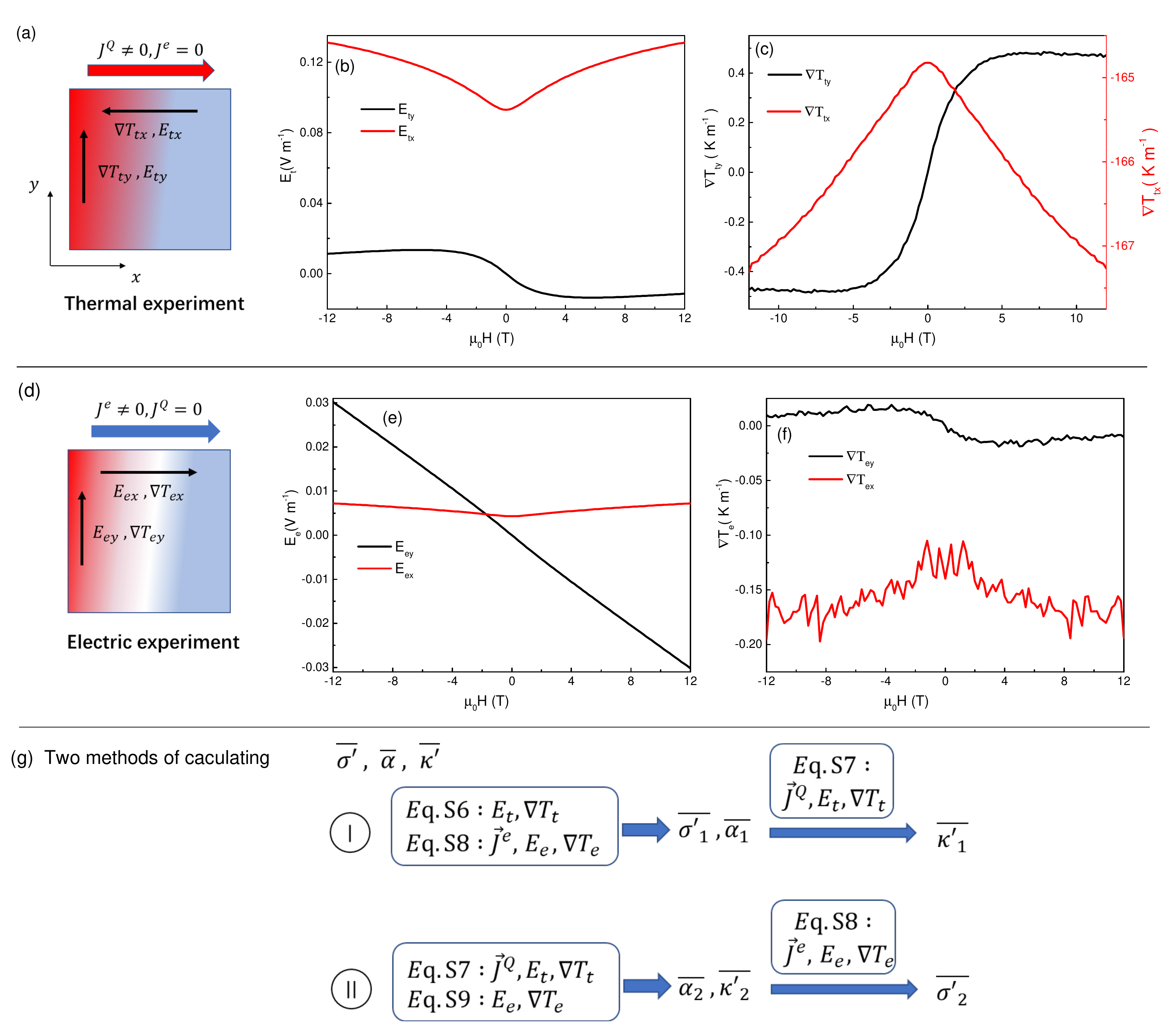}
    \caption{ (a)The first configuration of thermal experiment. (b) Electrical field and (c) thermal gradient measured in SrTiO$_{3-\delta}$ ($n$=1$\times$10$^{18}$ cm$^{-3}$) at $T$ = 19.4 K in the thermal experiment ($\Vec{J^Q}\neq\Vec{0}$ and $\Vec{J^e}$=$\Vec{0}$). (d)The second configuration of electric experiment. (e) and (f) same as (b) and (c) in the electric experiment ($\Vec{J^e}\neq0$ and $\Vec{J^Q}$=$\Vec{0}$).(g) Two methods by combining the two configurations to calculate $\overline{\sigma'}$, $\overline{\kappa'}$ and $\overline{\alpha}$. }
\label{fig: temperature difference and electric field}
\end{figure}

To determine the components of the 2 $\times$ 2 tensors $\overline{\sigma'}$, $\overline{\kappa'}$ and $\overline{\alpha} $, we conducted two types of experiments (See Fig. \ref{fig: temperature difference and electric field} a and d ). In the first one, a heat current is applied along the $x$-axis ($\Vec{J^Q}\neq\Vec{0}$) and electric current was kept at zero ($\Vec{J^e}$=$\Vec{0}$). In the second one, an electric current is applied along $x$ axis ($\Vec{J^e}\neq\Vec{0}$) without heat current ($\Vec{J^Q}$=$\Vec{0}$). In both experiments, we measured the temperature difference and the electric field along $x$ and $y$ axes at the same time. Fig. \ref{fig: temperature difference and electric field} b,c and f,g show typical data for both configurations at T= 19.4 K for SrTiO$_{3-\delta}$ ($n$=1$\times$10$^{18}$ cm$^{-3}$).

In the first configuration ($\Vec{J^Q}\neq0$ and $\Vec{J^e}$=$\Vec{0}$), Eq.\ref{S4} and Eq.\ref{S5} become:

\begin{equation}\label{S6}
\overline{\sigma'} \Vec{E_t} = \overline{\alpha} \Vec{\nabla T_t} 
\end{equation}

\begin{equation}\label{S7}
\Vec{J^Q}=\overline{\alpha} T \Vec{E_t}-\overline{\kappa'} \Vec{\nabla T_t} 
\end{equation}

In the second configuration ($\Vec{J^e}\neq0$ and $\Vec{J^Q}$=$\Vec{0}$) Eq.\ref{S4} and Eq.\ref{S5} become:

\begin{equation}\label{S8}
\Vec{J^e}=\overline{\sigma'} \Vec{E_e}-\overline{\alpha} \Vec{\nabla T_e} 
\end{equation}

\begin{equation}\label{S9}
\overline{\alpha} T \Vec{E_e}=\overline{\kappa'} \Vec{\nabla T_e} 
\end{equation}

\begin{figure}[h]
\centering
\includegraphics[width=17cm]{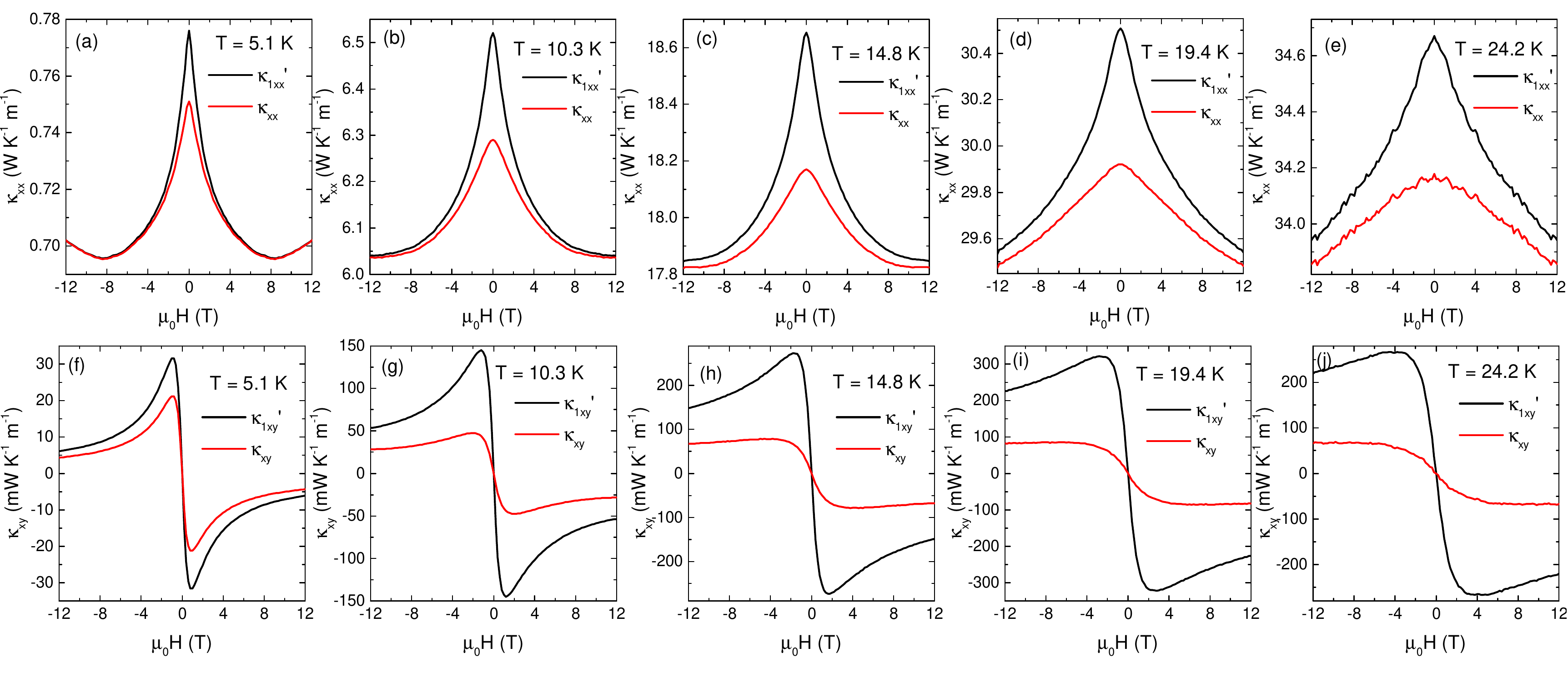}
    \caption{ (a)-(c) Longitudinal electric ($\sigma_{xx}$), thermoelectric ($\alpha_{xx}$) and thermal conductivity ($\kappa_{xx}$) as a function of field at T = 19.4 $K$ for SrTiO$_{3-\delta}$  ($n$ = 1$\times$10$^{18}$ cm$^{-3}$). The black and red colours correspond to two different combinations of Eq. \ref{S6}, \ref{S7}, \ref{S8} and \ref{S9}. Black (red) line results were obtained using Eq. \ref{S6}(\ref{S7}) \ref{S7}(\ref{S8}) and \ref{S8}(\ref{S9}). (d)-(f) same as (a)-(c) but for the transverse components of $\overline{\sigma^\prime}$, $\overline{\alpha}$ and $\overline{\kappa^\prime}$.}
\label{fig: kelvien relation}
\end{figure}

The four equations \ref{S6}, \ref{S7}, \ref{S8} and \ref{S9} projected along the $x$ and $y$ directions give us eight equations to determine six unknown quantities, which are the two components of the three tensors: $\overline{\sigma'}=\left[\begin{array}{ccc}
\sigma_{xx}' & \sigma_{xy}'\\
-\sigma_{xy}' & \sigma_{xx}'\\
\end{array}\right ] $, $\overline{\alpha}$= $\left[ \begin{array}{ccc}
\alpha_{xx} & \alpha_{xy}\\
-\alpha_{xy} & \alpha_{xx}\\
\end{array} 
\right ] $ and $\overline{\kappa'}$ = $\left[ \begin{array}{ccc}
\kappa_{xx}' & \kappa_{xy}'\\
-\kappa_{xy}' & \kappa_{xx}'\\
\end{array} 
\right ]$ assuming that they are isotropic in the xy plane. The redundancy allows us to make a check on self consistency. 

We calculated $\overline{\sigma'}$, $\overline{\alpha}$ and  $\overline{\kappa'}$ in two different ways, leading to $\overline{\kappa'}_{1,2}$ and $\overline{\sigma'}_{1,2}$, where the index refers to the set of equations and experimental data used (see Fig \ref{fig: temperature difference and electric field} (g)). Here, we give explicit expressions for deriving $\overline{\sigma'_1}$ and $\overline{\kappa'_1}$. 

Starting with Eq. \ref{S6} and \ref{S8} and setting $S = \sigma^{-1}\alpha$ , we obtain :

\begin{equation}\label{S10}
 S_{xx}=\frac{E_{tx}\nabla T_{tx}+E_{ty}\nabla T_{ty}}{\nabla T_{tx}^2+\nabla T_{ty}^2}; S_{xy}=\frac{E_{tx} \nabla T_{ty}-E_{ty}\nabla T_{tx}}{\nabla T_{tx}^2+\nabla T_{ty}^2}
\end{equation}

\begin{equation}\label{S11}
\sigma_{1xx}'^{-1}=\frac{E_{ex}-(S_{xx}\nabla T_{ex}+S_{xy}\nabla T_{ey})}{J^e};\sigma_{1xy}'^{-1}=\frac{E_{ey}-(-S_{xy}\nabla T_{ex}+S_{xx}\nabla T_{ey})}{-J^e}
\end{equation}

\begin{equation}\label{S12}
 \alpha_{1xx}=\sigma_{1xx}' S_{xx} - \sigma_{1xy}' S_{xy};  \alpha_{1xy}=\sigma_{1xx}' S_{xy} + \sigma_{1xy}' S_{xx}
\end{equation}

Injecting $\alpha$ into Eq. \ref{S8}, we get $\overline{\kappa'_1}$
\begin{equation}\label{S13}
 \kappa_{1xx}'=\frac{[T\cdot(\alpha_{1xx}E_{tx}+\alpha_{1xy}E_{ty})-J^Q]\cdot-\nabla T_{tx}-T\cdot(-\alpha_{1xy}E_{tx}+\alpha_{1xx}E_{ty})\cdot\nabla T_{ty}}{-\nabla T_{tx}^2-\nabla T_{ty}^2}
\end{equation}
\begin{equation}\label{S14}
 \kappa_{1xy}'=\frac{T\cdot(-\alpha_{1xy}E_{tx}+\alpha_{1xx}E_{ty})\cdot\nabla T_{tx}-[T\cdot(\alpha_{1xx}E_{tx}+\alpha_{1xy}E_{ty})-J^Q]\cdot-\nabla T_{ty}}{-\nabla T_{tx}^2-\nabla T_{ty}^2}
\end{equation}

 Method 2 starts with Eq. \ref{S7} and \ref{S9}  and leads to $\sigma'_2$, $\alpha_2$ and  $\kappa'_2$. Figure \ref{fig: kelvien relation} compares the results of the two methods. Onsager reciprocity implies no difference because of the identity between the components of the thermoelectric tensor in the two Eq.\ref{S4} and \ref{S5}. Experimentally, we find that this is indeed the case: $\sigma^\prime_{1ij}=\sigma^\prime_{2ij}$;  $\alpha^\prime_{1ij}=\alpha^\prime_{2ij}$ and $\kappa^\prime_{1ij}=\kappa^\prime_{2ij}$. Here, i and j refer to x and y orientations and 1 and 2 refer to the method 1 and method 2.

\begin{figure}[h]
\centering
\includegraphics[width=17cm]{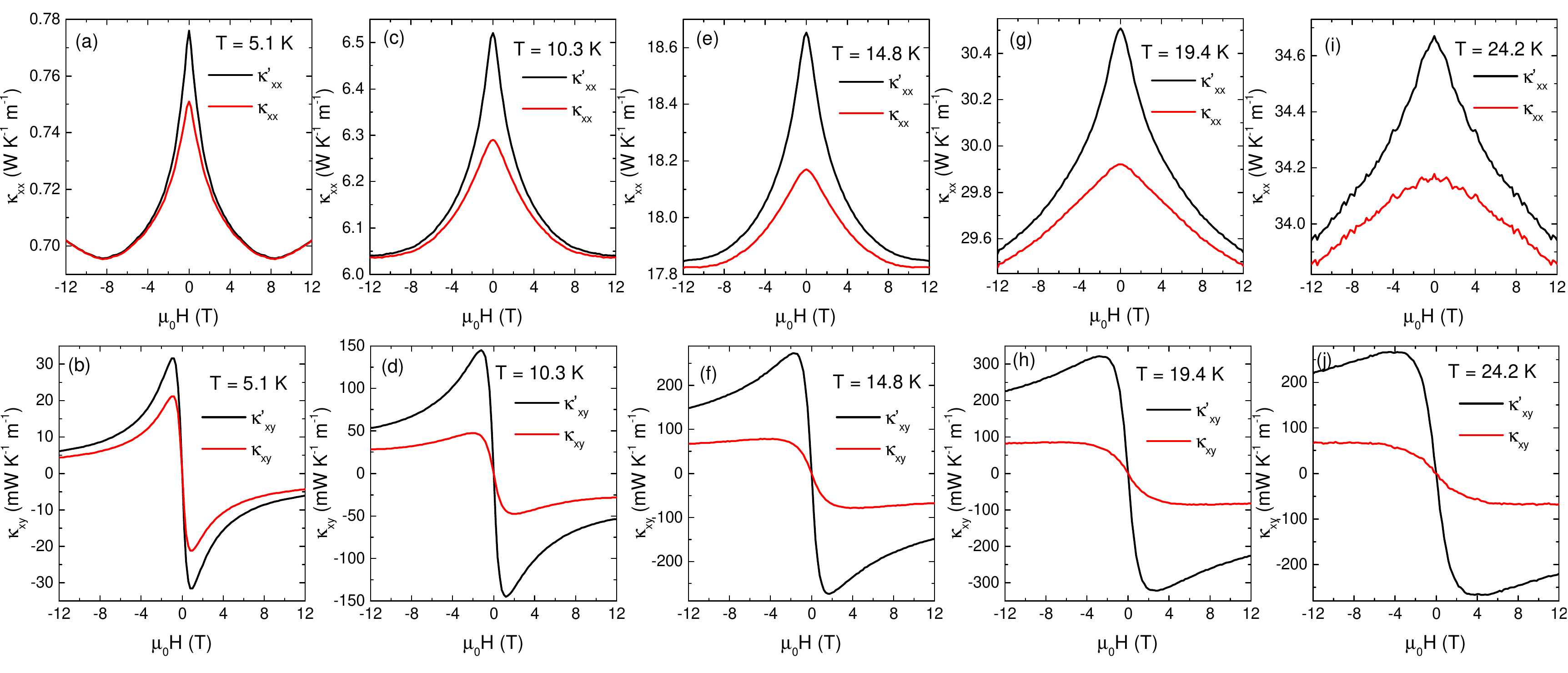}
    \caption{ (a)-(e) Field dependence of $\kappa_{xx}$ and $\kappa^\prime_{xx}$ for different temperature in SrTiO$_{3-\delta}$ ($n$=1$\times$10$^{18}$ cm$^{-3}$). (f)-(j) Same as (a)-(e) for $\kappa_{xy}$ and $\kappa^\prime_{xy}$.}
\label{fig: modification}
\end{figure}
\section{Quantifying the thermoelectric contribution to the thermal conductivity}
These measurements allowed  us to quantify the thermoelectric contribution to the thermal conductivity. Figure \ref{fig: modification} shows the difference between the components of $\overline{\kappa}$ and $\overline{\kappa^\prime}$ at different temperatures. The thermoelectric correction is tiny for the longitudinal thermal conductivity component: $\kappa^\prime_{xx}\approx\kappa_{xx}$ within a percent.  On the other hand, it is large for the transverse one: $\kappa^\prime_{xy}\neq \kappa_{xy}$.  In other words, the  flow of particles carrying entropy with no need for temperature gradient leads to a significant difference in the transverse channel. One can also see that the difference between $\kappa^\prime_{xy}$ and  $\kappa_{xy}$ decreases with temperature as discussed in the main text.

\end{document}